# The effect of diffusive re-equilibration time on trace element partitioning between alkali feldspar and trachytic melts

Arzilli F.[1], Fabbrizio A.[2], Schmidt M. W.[3], Petrelli M.[4], Maimaiti M.[5], Dingwell D. B.[6], Paris E.[5], Burton M.[1] and Carroll M. R.[5]

[1]School of Earth and Environmental Sciences, University of Manchester, Oxford Road, Manchester, M139PL, UK

[2]Institute of Petrology and Structural Geology, Faculty of Science, Charles University, Albertov 6, 12843 Prague, Czech Republic

[3]Institute for Mineralogy and Petrology, ETH Zurich, 8092 Zurich, Switzerland

[4]Deparment of Physics and Geology, University of Perugia, Piazza Università, Perugia 06100, Italy

[5]School of Science and Technology, Geology Division, University of Camerino, Via Gentile III da Varano, Camerino, 62032, Italy

[6]Earth and Environmental Sciences, LMU Munich, Theresienstrasse 41/III, Munich 80333, Germany

*Corresponding author: Dr. Fabio Arzilli, PhD

Corresponding author affiliation: School of Earth and Environmental Sciences, University of Manchester, Oxford Road, Manchester, M139PL, UK

E-mail: fabio.arzilli@manchester.ac.uk; arzilli.fabio@gmail.com

Keywords: Alkali feldspar, Trachytic melts, Trace elements, Partition coefficient, Disequilibrium crystal growth, Diffusive re-equilibration

**Abstract**

We present new experimental data on major and trace element partition coefficients between alkali feldspar and trachytic melt. Experiments were conducted at 500 MPa, 870-890 °C to investigate through short disequilibrium and long near-equilibrium experiments the influence of diffusive re-equilibration on trace element partitioning during crystallization. Our data show that Ba and Sr behave compatibly, and their partition coefficients are influenced by re-equilibration time, orthoclase (Or) content, growth rate and cation order-disorder. High field strength elements (HFSE) and rare earth elements (except Eu) are strongly incompatible, but alkali feldspar efficiently fractionates light (LREE) from heavy rare earth elements (HREE).

Our crystallization experiments reveal a strong influence of disequilibrium crystal growth on the partitioning of Ba and Sr. In particular, short-duration experiments show that rapid alkali feldspar crystal growth after nucleation, promotes disordered growth and less selectivity in the partitioning of compatible trace elements that easily enter the crystal lattice (e.g., Ba and Sr). This produces partition coefficients of compatible elements higher than those obtained through long-duration experiments, in which growth is slower and more selective. On approach to equilibrium, decreasing growth rate and timescales of diffusive re-equilibration facilitates the incorporation of compatible elements within the crystal lattice (through ordered growth), on the basis of the charge and size of the octahedral site, resulting in lower partition coefficients of Ba and Sr. Therefore, our results indicate that besides crystal-chemical effects, i.e. lattice strain, the substitution mechanisms during growth (with different degrees of ordering) of Ba and Sr in alkali feldspar are strongly influenced by diffusive re-equilibration and crystal growth kinetics. This implies that partition coefficients of Ba and Sr can be used to establish timescales of crystallization processes under pre- and syn- eruptive conditions. The application of our results to alkali feldspar in rocks from Campi Flegrei, constrain the magma residence time at subliquidus conditions in a reservoir to a maximum of 6 days under disequilibrium conditions and to a minimum of 9 days upon approaching near-equilibrium conditions.

1. Introduction

Trace element geochemistry plays a fundamental role in understanding the evolution of the Earth. The chemical analyses of trace elements and their isotope ratios in rocks, minerals and melts (glass) can hence provide quantitative evidence concerning the origin and subsequent differentiation of the Earth (Blundy and Wood, 2003).

The partition coefficient of an *element i* ($D_i$) has been defined as the concentration of an element in a mineral divided by the concentration of that element in the coexisting melt (Blundy and Wood, 1994; Jones, 1995) and describes the chemical distribution between coexisting phases under equilibrium conditions. Disequilibrium conditions however can also affect partitioning and thus partition coefficients (Albarede and Bottinga, 1972). Here we distinguish such "apparent" partition coefficients related to kinetic effects from equilibrium partitioning coefficients (Blundy and Wood, 1994; Jones, 1995). If Henry's Law is satisfied, partition coefficients will be independent of the concentration of the trace element. Therefore, apparent $D_i$ can be used as sensitive monitors of magmatic processes such as fractional crystallization, assimilation, mixing and partial melting of crustal or mantle rocks (Drexler et al., 1983; Henderson and Pierozyski, 2012; Leeman and Phelps, 1981; Mahood and Hildreth, 1983; White et al., 2003), for which trace element abundances can vary by orders of magnitude, while major elements may vary only by a few percent. Furthermore, partition coefficients can vary significantly as they are complex functions of crystal structure and composition, melt structure and composition, temperature, pressure, oxygen fugacity, and kinetic effects (Aigner-Torres et al., 2007; Drake and Weill, 1975; Icenhower and London, 1996; Irving, 1978; Long, 1978;

Mahood and Stimac, 1990; Morgan and London, 2003; Mollo et al., 2011; Schmidt et al., 2006; Stix and Gorton, 1990; Sun et al., 2017; Van Westrenen et al., 1999; White et al., 2003).

In natural or laboratory systems, crystallization can involve both disequilibrium and equilibrium processes, and the texture and chemical composition of crystals typically record their growth history. Most partition experiments are applied at equilibrium conditions, assuming that natural systems are characterized by chemical homogeneity of melt and crystals and that these systems evolve slowly enough to allow for continuous re-equilibration between melt and growing crystals (Higuchi and Nagasawa, 1969; Onuma et al., 1968; Schnetzler and Philpotts, 1970; Philpotts and Schnetzler, 1970). The disequilibrium partitioning of trace elements between crystals and coexisting silicate melt was introduced to the geological community by Albarede and Bottinga (1972), who modelled and documented a correlation between grain size and trace-element zoning patterns in igneous rocks.

The regimes of crystal growth in igneous systems have been classified as interface-controlled and diffusion-controlled (e.g., Dowty 1980; Hammer 2008; Kirkpatrick 1981; Lofgren 1980). Crystallization is interface-controlled when the attachment of an element on a nucleus or on crystals is slower than its diffusion in the melt. In contrast, diffusion-controlled crystallization is the opposite situation. In general, these two situations reflect different degrees of undercooling, as the former process is favoured at low undercooling (near-equilibrium conditions), whereas the latter prevails at high undercooling and disequilibrium conditions. Nevertheless, the crystallization regime also depend on the element diffusivity. Several studies show that in some natural igneous systems, rapid crystal growth can produce a disequilibrium accumulation of elements near the growing mineral interface; this results in a diffusive boundary layer in the growth medium surrounding the crystal, and consequent disequilibrium element uptake during crystal growth (e.g., Albarede and Bottinga, 1972; Baker, 2008; Mollo et al., 2011; Mollo et al., 2013; Morgan and London, 2003; Watson, 1996;

Watson and Muller, 2009). During progressive crystal growth, element fractionation occurs within this boundary layer where diffusion rates of individual elements vary with size and charge (Baker, 2008; Blundy and Wood, 1994; Blundy and Wood, 2003; Watson and Muller, 2009). During rapid crystal growth nutrients are supplied in different proportions than under equilibrium conditions (e.g., Baker, 2008; Lofgren et al. 2006; Mollo et al., 2011; Mollo et al., 2013; Morgan and London, 2003) and trace elements partitioning can be controlled by crystallization kinetics.

Quantitative experimental data on the role of crystallization kinetics on trace elements partitioning relative to high-pressure conditions and relevant for deep magmatic reservoir are still scarce. Yet, these would be especially important for alkali feldspar, as it is the volumetrically dominant crystal phase in many evolved magmas such as trachytes, phonolites, pantellerites and rhyolites. Its abundance may vary strongly with small changes in pressure, temperature and water content (e.g., Arzilli and Carroll, 2013), implying appreciable variations in textures and magma rheology. Furthermore, alkali feldspar in evolved magmas dominates bulk partition coefficient for many trace elements during fractional crystallization (Mahood and Hildreth 1983; Mahood and Stimac, 1990; Morgan and London, 2003; Nicholls and Carmichael, 1969; Roux and Varet, 1975; White et al., 2003). Therefore, it is fundamental to quantitatively investigate kinetic effects during crystal growth on trace element partitioning.

To investigate these effects, we conducted short- and long-duration experiments on alkali-feldspar in trachytic melt, simulating both disequilibrium and near-equilibrium crystallization conditions. Relationships between ionic radius, trace element valence and partition coefficients then demonstrate the differences for crystals grown in short disequilibrium vs. long equilibrium experiments. This study has strong implications for natural magmatic systems that can rapidly pass from equilibrium to disequilibrium conditions (e.g., magma mixing, pre-eruptive magma ascent) and it allows us to better understand crystallization kinetics and magma differentiation in crustal magma chambers.

## 2. Experimental and analytical methods

### 2.1 Starting materials

The starting material is a trachytic obsidian (ZAC; Table 1) from the basal unit of the Breccia Museo Member of the Campanian Ignimbrite eruption (39 ka; Fedele et al. 2008; Melluso et al. 1995). The obsidian is anhydrous and crystal-poor (<10 vol.%). Mineral phases are mostly alkali feldspar, clinopyroxene, biotite and magnetite, in order of decreasing abundance (Fabbrizio and Carroll 2008; Fedele et al. 2008). The ZAC composition was chosen because its phase relations are well known (Arzilli and Carroll 2013; Fabbrizio and Carroll 2008). Fragments of the obsidian were finely ground in an agate mortar with ethanol to a mesh size of about 20 μm. The obsidian was twice powdered and fused for 1 hour at 1300 °C (at ambient pressure) in a Pt crucible, in order to homogenise the melt and destroy all nuclei. The resulting glass was reground with ethanol and about 1 wt.% of a mix of trace elements (TX7 powder; Supplementary Table 1) was added and then reground again with ethanol.

### 2.2 Experimental strategy

Experiments with duration ranging from 6 to 192 hours were performed at 500 MPa and water undersaturated conditions (~2 wt. % of $H_2O$) at 870-890 °C, in order to induce the crystallization of alkali feldspar. The liquidus of alkali feldspar, with 2 wt. % $H_2O$ dissolved in the melt, has a positive slope of 10 °C/100 MPa (Fig. 1) as estimated from Burnham (1979). The liquidus of alkali feldspar at water-undersaturated conditions was constrained using the water-saturated ZAC phase diagram (Arzilli and Carroll, 2013; Fabbrizio and Carroll, 2008) and verified through the experiments performed in this study.

Isobaric cooling experiments with short duration (18-39 hours) were performed using an end-loaded piston cylinder apparatus (3/4 inch) at the Institute of Geochemistry and Petrology, ETH Zurich (Switzerland). About 0.8 µl of deionized water were added with a micro-syringe in $Au_{80}Pd_{20}$ capsules (6 mm length, 3.5 mm internal diameter, 0.25 mm wall thickness), then 40 mg of the starting material were added to reach the desired water content of ~2 wt. %. Capsules were welded shut, weighed, immersed for 10 min in acetone and finally re-weighed to verify the absence of leaks. The cylindrical assemblies consisted of a lead foil wrapped around a talc sleeve, a Pyrex sleeve, a graphite furnace and inner pieces of crushable magnesia. Experimental *P-T* conditions were chosen using the phase diagram reported in Fig. 1. Temperatures were measured using type B $Pt_{94}Rh_6-Pt_{70}Rh_{30}$ thermocouples. Experiments were initially held for 6 hours above the liquidus at 500 MPa and 1100 °C ($T_i$) (Fig. 1). Then temperature was rapidly decreased (50 °C/min) to the final temperature ($T_f$ = 890, 885, 880 and 870 °C) and held for 18-39 hours in order to induce the crystallization of alkali feldspar. At the end, the experiments were quenched by switching off power to the furnace.

Long duration experiments of 8 days were performed at Laboratoire Magmas et Volcans, Blaise Pascal University, Clermont-Ferrand (France), using a non end-loaded piston-cylinder apparatus (3/4 inch). The starting material was contained in Au capsules with 10 mm length, an outer diameter of 4 mm and 0.25 mm wall thickness. The capsules were loaded with deionized water and about 32 mg of the starting powder (reaching 2 wt. % $H_2O$), then welded, squeezed to a cylindrical shape with a length of 6-7 mm, and tested with acetone in order to insure that they were sealed. The 3/4 inch piston-cylinder assembly consisted of a Teflon foil wrapped around a salt sleeve, a Pyrex sleeve, a graphite furnace, an inner Pyrex sleeve, and inner pieces of crushable magnesia. Temperatures were measured with type C thermocouples (WRe$_5$/WRe$_{26}$). The experiments were carried out at 500 MPa and initially held at 1040 °C for 6 hours to dissolve potential crystal nuclei before decreasing the temperature at 50 °C/min to 880 °C where they were held for 192 hours and then quenched to <400 °C within 10 s maintaining pressure.

The liquidus temperature of alkali feldspar at 500 MPa and 2 wt. % of H$_2$O is ~920 °C (Fig. 1). Thus the degree of undercooling (with respect to the liquidus of alkali feldspar) in our experiments ranges between ~ 30 and ~50 °C. Experimental conditions are summarized in Table 2.

*2.3 Analytical methods*

*2.3.1 Textural analysis*

Back-scattered electron (BSE) images were collected using a JEOL JSM-6390LA field emission-scanning electron microscope (SEM) at ETH Zurich and a JEOL 5910LV scanning electron microscope at LMV, Clermont-Ferrand. Analytical conditions were 15 kV accelerating voltage and 20 nA beam current.

BSE images were used to investigate sample textures and to estimate the growth rate of alkali feldspar. The sizes of alkali feldspar crystals were measured through BSE images, using ImageJ software (NIH Image; Abramoff et al., 2004; Schneider et al. 2012). The largest 10 crystals in each sample were measured (e.g., Arzilli and Carroll, 2013; Couch, 2003) to estimate the maximum growth rate ($Y_L$). The growth rate was calculated using only the longest dimension (L) of each crystal and the duration (*t*) of the experiment (e.g., Arzilli and Carroll, 2013; Couch, 2003; Couch et al., 2003; Fenn 1977; Shea and Hammer 2013; Swanson, 1977): $Y_L = (L * 0.5)/t$.

*2.3.2 Electron microprobe analysis (EMPA)*

Major elements were analyzed using a JEOL JXA-8200 Electron Microprobe Analyzer (EMPA) at ETH Zurich. Glass and crystal compositions were analyzed using 15 kV accelerating voltage, 10 nA beam current and a beam size of 10 μm for glass or 5 μm for crystals. Standards for calibration were quartz for Si, corundum for Al, rutile for Ti, periclase for Mg, pyrolusite for Mn, aegirine for Na, sanidine for K, wollastonite for Ca. Sodium and potassium were measured first to minimize their migration. Measurements times were 10 s on peak and 5 s on background.

Chemical analyses of equilibrium experiments T11 and T12 were obtained with an electron microprobe Cameca SX 50 at CNR Institute for Geosciences and Earth Resources, University of Padova (Italy). Analytical conditions were similar to those used at ETH Zürich.

*2.3.3 Laser Ablation - Inductively Coupled Plasma Mass Spectrometry (LA-ICP-MS)*

LA-ICP-MS analyses of samples T2, T6, T8, T9 and T10 were performed at ETH Zurich using an ArF 193 nm excimer laser ablation system (Gunther et al., 1997) combined with a Perkin-Elmer Elan DRC 6000 ICP- MS instrument. The maximum output energy of the laser was 200 mJ per pulse, the repetition rate 10 Hz and the beam diameter 40 μm.

Trace element concentrations for samples T11 and T12 were determined at the Department of Physics and Geology, University of Perugia (Italy). The instrumentation consisted of a Teledyne/Photon Machine G2 LA device equipped with a Two-Volume ANU HelEx 2 cell coupled with a Thermo Scientific quadrupole-based iCAP Q ICP-MS. Analyses of alkali feldspar crystals and trachytic glasses were performed by using a circular laser beam of 20 μm diameter, a frequency of 8 Hz and a laser density on the sample surface of 3.5 J/cm$^2$. USGS BCR2G reference material was analyzed as unknown in order to provide a quality control (Wilson, 1997). For both systems, the NIST SRM-610 reference glass was used as calibrant and Si, previously analysed by EPMA, as internal standard. Under these operating conditions precision and accuracy are better than 10 % for all elements (Gunther et al., 1997; Petrelli et al., 2007, 2008, 2016).

The acquisition of time resolved signals allows the identification of melt inclusions in the crystals. Thus, the raw data were carefully screened for such inclusion signals, which were eventually removed.

### 3. Results

*3.1 Textural features and growth rates of alkali feldspar*

Mineral phases present in the short-term experiments are alkali feldspar and <5 vol.% clinopyroxene (Fig. 2a-b). Both euhedral and anhedral alkali feldspars were crystallized. Different crystal morphologies occurred, such as tabular, prismatic elongated, acicular and skeletal shapes (Fig. 2a-b). Most crystals show hopper morphologies with the crystal length more developed in comparison with the other two directions, showing that the formation of hopper shapes is related to rapid growth (Iezzi et al., 2014; Lofgren, 1974; Muncill and Lasaga 1987; 1988; Vetere et al., 2013; 2015; Vona and Romano, 2013). Growth rates ($Y_L$) are in the range $1.4\text{-}5.2\times10^{-7}$ cm/s (Table 3), in agreement with Arzilli and Carroll (2013).

Long-duration experiments (T11 and T12) are characterized by alkali feldspar and clinopyroxene (Fig. 2c-d). Textures show mostly euhedral crystals of alkali feldspar characterized by tabular and prismatic elongated shapes (Fig. 2c-d). The apparent growth rates ($Y_L$) of alkali feldspar are between $5.4\times10^{-9}$ and $1.7\times10^{-8}$ cm/s (Table 3).

*3.2 Partition coefficients between alkali feldspar and trachytic melts*

Major and trace element analyses of alkali feldspar crystals and trachytic glasses are listed in Table 4 and 5, and partition coefficients are summarized in Table 6. In the following, we present results obtained from short and long duration experiments, respectively.

*3.2.1 Short-duration experiments*

*Monovalent cations*: monovalent cations are compatible (K) and incompatible (Na, Rb, Li and Cs) (Table 6). The distribution coefficient for potassium ($D_K$) is always higher than unity, $D_{Na}$ is 0.6-0.7 $D_{Rb}$ is 0.3-0.4, and $D_{Li}$ and $D_{Cs}$ are <0.1.

*Divalent cations*: Ba and Sr are highly compatible in alkali feldspar, with $D_{Ba}$ ranging from 27 to 64 and $D_{Sr}$ from 16 to 30 (Table 6). Instead, Ca is incompatible ($D_{Ca}$ = 0.60-0.95). Other divalent cations such as Be, Zn and Co are incompatible, with partition coefficients <0.1.

*Trivalent REE*: most of the rare earth elements are strongly incompatible in alkali feldspar (Table 6), only Eu shows a more complex behaviour ranging from incompatible to compatible (~0.7<$D_{Eu}$<1.4), probably due to variations of its oxidation state. Distribution coefficients show that LREE (La, Ce, Pr and Nd) are slightly less incompatible in alkali feldspar than MREE (Sm, Gd, Tb and Dy) and HREE (Er, Yb and Lu), $D_Y$ ranges from 0.002-0.005.

*Other cations*: High Field Strength Elements (HFSE) are highly incompatibles in alkali feldspar, with distribution coefficient values for Zr, Nb, Hf, Ta, Th and U lower than 0.06 (Table 6).

*3.2.2 Long-duration experiments*

*Monovalent cations*: K remains compatible, the distribution coefficients for Na and Rb range from 0.2-0.7, showing that both are incompatible, $D_{Li}$ and $D_{Cs}$ remain <0.1 (Table 6).

*Divalent cations:* Barium and Strontium are still highly compatible in alkali feldspar but distribution coefficients of 4-16 are significantly lower than in the short experiments. Ca is again incompatible (~0.25<$D_{Ca}$<0.45).

*Trivalent REE*: Rare earth elements (La, Ce, Pr, Nd, Tb) and Eu remain incompatible in alkali feldspar (Table 6).

*Other cations*: HFS elements are highly incompatible in alkali feldspar with distribution coefficients lower than 0.04 (Table 6).

## 4. Discussion

### *4.1 Site occupancy*

The general formula of alkali feldspar is $AT_4O_8$, where A represents the X-fold coordinated site hosting alkalis, the alkaline earths and other cations such as $Fe^{2+}$ and $Mn^{2+}$ and all U-series elements. T is the IV-fold coordinated site hosting Ti and $Fe^{3+}$ replacing Al (Blundy and Wood 2003; Deer et al., 1967; Fabbrizio et al., 2009). However, U, Th and Nb have much higher charges and smaller ionic radii than K, their ability to occupy the octahedral site is hence limited compared to mono- and divalent ions, consequently, they behave extremely incompatibly and are characterized by low distribution coefficients (Blundy and Wood, 2003). The similarity in ionic radius of K, Ba and Sr and the relatively flexible environment of the A-site in alkali feldspar (Blundy and Wood, 2003; Fabbrizio et al., 2009) allows Ba and somewhat less Sr to be easily hosted in alkali-feldspar (Ewart and Griffin, 1994; Fabbrizio et al., 2009; Icenhower and London, 1996); charge balance being maintained by appropriate replacement of Si by Al (Guo and Green, 1989; Icenhower and London, 1996; Long, 1978). Sr is less favoured than Ba (1.52 Å) as the former causes a greater distortion of the crystal lattice due to its ionic radius (1.36 Å) being smaller than the ideal size of the X-fold coordinated A-site in alkali feldspar (1.462-1.468 Å).

The differences between the ionic radius of K and those of Rb, Cs and Li render the latter three incompatible within alkali feldspar structures (Fabbrizio et al., 2009; Henderson and Pierozynski, 2012; Icenhower and London, 1996; Long, 1978; Morgan and London, 2003).

Trivalent cations such as the REEs are strongly incompatible in alkali feldspar, as their valence produces a charge imbalance in the octahedral site. However, La, Ce, Eu, Pr and Nd can be present in significant amounts (Leeman and Phelps, 1981) because their ionic radius is similar to that of Na.

*4.2 Lattice strain model (LSM)*

Trace-element distribution coefficients ($D_i$) in alkali feldspar are influenced by crystal chemistry and melt composition (Blundy and Wood, 1994; Higuchi and Nagasawa, 1969; Icenhower and London, 1996; Nash and Crecraft, 1985; Ren, 2004; Schmidt et al., 2006). In addition to ionic size, elastic properties of the mineral lattice may also influence elemental diffusion and, thus, the distribution coefficient (Blundy and Wood, 1994; Cherniak, 2002). The partitioning behaviour for cations ($D_i$) with the same valence is conveniently shown in Onuma diagrams as a function of ionic radius (Blundy and Wood, 1994; Leeman and Phelps, 1981; Onuma et al., 1968) (Figs 3 and 4), which indicates why given elements are compatibles in a given site.

The relationship between the distribution coefficient of element $i$ and its ionic radius is described by the following equation (Brice, 1975; Blundy and Wood, 1994):

$$\log D_i = \log D_0 + 0.43429 \left[ \frac{-91.017 \, E \left[ \frac{r_0}{2}(r_i - r_0)^2 + \frac{1}{3}(r_i - r_0)^3 \right]}{T} \right] \quad \text{Eq. 1}$$

where $D_0$ is the strain-compensated partition coefficient for a fictive cation with radius $r_0$ that would fit into the crystallographic site without strain, $E$ is the Young's Modulus of the concerned site, $r_i$ is the radius of the cation of interest, and $T$ is temperature in Kelvin. This model is the mathematical description of an Onuma parabola (Onuma et al., 1968). The distribution coefficients of selected mono- and divalent cations, fitted using Eq. (1) are summarized in Figs 3 and 4, and listed in Table 7.

*4.2.1 Onuma diagrams for short duration experiments*

Onuma diagrams (Fig. 3) show that element partitioning in alkali feldspar in our short experiments is always similar within the narrow temperature range investigated (870-890°C). Monovalent cations show single peak curves with the highest $D_i$ near K, and the calculated ideal site size $r_0$ or maximum of the inverted parabola is at 1.462-1.468 Å, similar to the ionic radius of K (1.59 Å) (Fig. 3, Table 7) (Fabbrizio et al., 2009). Parabolas for divalent cations show Ba close to the parabolas' apex, located at $r_0 = 1.470$ Å (run T9), confirming that Ba can easily substitute for K in the A-site. The increased lattice strain due to the ionic radius difference between K and other elements such as Na, Rb, Cs, and Ca results in lower partition coefficients relative to Ba and Sr (e.g.: Blundy and Wood, 1994; Ren, 2004). As to be expected, divalent cations describe tighter parabolas relative to the monovalent ones, yielding higher Young's moduli (Table 7).

Values of $D_0$, $r_0$ and $E_0$ for mono- and divalent cations as fitted to the experimental data are given in Table 7, together with $E_c$ and $r_c$ calculated following equations from Blundy and Wood (2003):

$$r_c^2 = 1.341 + 0.207 X_{Or} \text{ (Å)} \qquad \text{Eq. 2}$$

$$E_c = 1125 \quad valence \quad \left(r_c^{n+} + 1.38\right)^3 \text{ (GPa)} \qquad \text{Eq. 3}$$

Trivalent cations are all incompatible, showing the highest distribution coefficient for La and decreasing from La to Lu. Because the LREEs (La through Nd) are more compatible than HREE, alkali feldspar is slightly enriched in LREE (Fabbrizio et al., 2009; Mahood and Hildreth, 1983; Mahood and Stimac, 1990; Ren, 2004; Stix and Gorton, 1990; White et al., 2003).

Tetravalent cations are strongly incompatible in alkali feldspar because of their valence and radius differences to the A-site (Fig. 3; Table 6), suggesting that their substitution involves structural defects.

*4.2.2 Onuma diagrams for long experiments*

Onuma diagrams for the longer duration experiments are qualitatively similar to those of the short runs. Monovalent cations show single peak curves with the highest $D_i$ values at $r_0$ of 1.452 Å (run T11) and 1.476 Å (run T12) (Fig. 4), corresponding to the ideal size of the X-fold coordinated A-site in alkali feldspar (Fabbrizio et al., 2009). Divalent cations show Ba close to the maximum of the parabolas, (Fig. 4) the radius of Ba fitting the A-site well and causing the least lattice strain. Ba is again more compatible than Sr because of its proximity in ionic radius to K. An important finding from the long duration experiments is that the partition coefficients of Ba and Sr, and in lesser extent those of Ca, are significantly lower than those obtained from short duration experiments but partition coefficients of monovalent cations do not show substantial differences between the two series. Distribution coefficients of LREE (La, Ce, Pr and Nd) from long-duration experiments are up to one order of magnitude lower than those of short-duration experiments (Figs 3 and 4; Table 6). These observations indicate that the experimental time affects trace element partitioning, influencing the $D_i$ for divalent and trivalent trace cations.

*4.3 Comparison with synthetic systems*

Several experimental studies have investigated the partitioning behaviour of mono- and divalent elements between alkali feldspar and silicate melts over a range of synthetic systems approximating natural rhyolite (Henderson and Pierozynski, 2012; Long 1978; Morgan and London, 2003), metapelite (Icenhower and London, 1996), trachyte (Fabbrizio et al. 2009; Guo and Green, 1989;

Henderson and Pierozynski, 2012), and phonolite (Henderson and Pierozynski, 2012). Our results are broadly in agreement with these previous studies, but our new data also provide insight into how kinetic factors may influence crystal-melt partitioning behaviour.

We observe that Ba and Sr always result compatible in alkali feldspar, whereas Rb and Cs are incompatible. In contrast, the behaviour of Na, K and Ca depends on the chemical composition of alkali-feldspar and on experimental durations, i.e. kinetic vs equilibrium conditions. In addition, our results for short and long duration experiments show that $D_{Ba}$, $D_{Sr}$ and $D_{Rb}$ increase with increasing orthoclase content (Fig. 5a, b, c), a trend also reported from syntethic granitic, trachytic and peraluminous melts (Guo and Green, 1989; Henderson and Pierozynski, 2012; Icenhower and London, 1996; Long, 1978).

Given the wide range in distribution coefficients for Ba and Sr and experimental times ranging from a few hours to 8 days, it is reasonable to ask whether all experimental distribution coefficients represent equilibrium partitioning between alkali feldspar and coexisting glass. Consequently, partition coefficients from previous experimental works were re-fitted using Eq. 1. Then, these partition coefficients were filtered for plotting on their respective parabola, in addition only those were accepted which parabola yields $E^{2+}$ values in the range 55-160 GPa (Blundy and Wood, 2003). The filtered distribution coefficients are reported in Table 8, and compared with our data in Fig. 6a. In general, there is good agreement between our distribution coefficients and those retained for different melt compositions (Fabbrizio et al. 2009; Henderson and Pierozynski, 2012; Icenhower and London, 2006; Morgan and London, 2003) following the above quality control procedure. This suggests that for the strongly polymerized melts treated here, melt polymerization (i.e. NBO/T ratio) does not play a major role in controlling alkali-feldspar/silicate liquid mono- and divalent cations partitioning (in agreement with White et al., 2003). Furthermore, our $D_{Ba}$ values from the short experiments (less than 40 hours) deviate most from the selected data, whereas those from the long runs (8 days) agree well with literature data obtained with run durations of at least 5 days (Fabbrizio et al. 2009; Henderson and Pierozynski, 2012; Icenhower and London, 2006; Morgan and London, 2003). This implies that run duration is a key factor in controlling the partitioning of Ba, and that excessively short runs can lead to an overestimation of the Ba equilibrium distribution coefficient.

*4.4 Comparison with natural systems*

To obtain a reliable comparison between distribution coefficients derived experimentally and from natural rocks, we selected only literature data obtained from pairs of coexisting homogeneous alkali-feldspar and unaltered glass in trachytic samples. The selected data set (Table 9) comprises 15 trachytic and trachy-phonolitic samples from Campi Flegrei, Italy (Fedele et al., 2015; Pappalardo et al., 2008; Villemant, 1988), from Pantelleria, Italy (White et al., 2003), and from Baitoushan, China-North Korea (White et al., 2003). These apparent alkali-feldspar/liquid distribution coefficients show few differences with the results of our long-duration experiments (Fig. 6b and Table 9). Instead, apparent partition coefficients, related to our short-duration experiments, are much higher than those of natural systems. Our $D_{Ba}$ values of 7.8 and 15.6, obtained from long-duration experiments (samples T11 and T12), replicate $D_{Ba}$ values of 7.7 and 16.6 for trachytic and trachyphonolitic samples from Campi Flegrei (Fedele et al., 2015; Pappalardo et al., 2008; Villemant, 1988), our Or contents range between 32 and 57 mol.%, while Campi Flegrei sanidine crystals have 57.9-62.9 mol.% Or. Our $D_{Ba}$ values are also in the range of values (1.9-12.2) determined for trachyte from Pantelleria (White et al., 2003), in which sanidine has 30.9-36 mol.% Or. Our $D_{Sr}$ values of 4.1 and 9.6 compare well with the variation range ($D_{Sr}$ = 2-8.8) in trachyte and trachyphonolite from Campi Flegrei (Fedele et al., 2015; Pappalardo et al., 2008; Villemant, 1988) as well as with a $D_{Sr}$ of 4 determined for the Baitoushan trachyte (White et al. 2003).

Trachytes from Pantelleria (White et al., 2003) show $D_{Rb}$ values (0.2-0.3) close to our data (0.2-0.5), whereas those from Campi Flegrei (0.7-0.97) are slightly higher (Fedele et al., 2015; Pappalardo et al., 2008; Villemant, 1988). Although Rb is incompatible, its partition coefficient slightly increases as Or contents increase (Fig. 5c), due to the geochemical similarity in terms of charge and size of Rb and K (Icenhower and London, 1996; Mahood and Stimac, 1990; Ren, 2004; White et al., 2003). However, no significant difference is noted between the $D_{Rb}$ values (Table 6) calculated from short- and long-duration experiments and most of the literature (Drexler et al., 1983; Fedele et al., 2015; Icenhower and London, 1996; Leeman and Phelps, 1981; Mahood and Hildreth, 1983; Mahood and Stimac, 1990; Pappalardo et al., 2008; Ren, 2004; Villemant 1988; White et al., 2003), suggesting that in contrast to Ba and Sr the partitioning of Rb is not very sensitive to changes in melt composition or crystallization conditions or kinetics.

Our $D_{Cs}$ values (0.005-0.04) are comparable to those obtained for trachyte (0.05-0.07) by Fedele et al. (2015), whereas those from Villemant (1988) are sligthly higher ($D_{Cs}$ = 0.08-0.12). Experimental values of $D_{Na}$ (0.52-0.73) and $D_K$ (1.1-1.6) match well with a $D_{Na}$ (0.44-0.55) and $D_K$ (1.5-1.6) for trachyte from C ampi Flegrei (Fedele et al., 2015). REE distribution coefficients calculated from our data do not show significant differences with those for trachyte from Campi Flegrei and Pantelleria (Fedele et al., 2015; Pappalardo et al., 2008; Villemant, 1988; White et al., 2003).

*4.5 Influence of time on the element partitioning: disequilibrium vs. equilibrium conditions*

In alkali feldspars, the sum of monovalent cations (K, Na, Rb, and Cs) + silica vs. divalent cations (Ca, Sr, and Ba) + Al (Fig. 7a) is close to the line described by the equation $M^{1+} + Si^{4+} = - (M^{2+} + Al^{3+}) + 5$. This suggests that the alkali feldspar compositions are nearly stoichiometric, divalent cations in the alkali feldspar lattice being accommodated by the exchange reaction:

$M^{1+} + Si^{4+} = M^{2+} + Al^{3+}$                                                                                                          Eq. 4

In general we note that high partition coefficients of divalent cations are related to high Or contents (Fig. 5a, b, d). However, alkali feldspar crystals with similar Or contents show different $D_{Ba}$, $D_{Sr}$ and $D_{Ca}$ (Fig. 5) and different concentrations in alkali feldspar (Fig. 7b). Our results show that the concentrations of divalent elements and their partition coefficients decrease as experimental times increase (Fig. 7c, d respectively). This observation implies that alkali feldspar trace element compositions, grown from a given bulk composition, depend on the time experienced at subliquidus conditions. This means that the initial rapid growth is followed by a time of diffusive equilibration, which then brings the alkali feldspar to the equilibrium composition. Hence, crystallization time is able to affect the partition coefficient of Ba and Sr between alkali-feldspar and silicate liquid, with high values resulting from short-duration experiments (18-39 hours), while long near-equilibrium experiments yield significantly lower partition coefficients (Figs 5a, b, d and 7d). This process can explain high (e.g., Ewart and Griffin, 1984; Fedele et al., 2015; Leeman and Phelps, 1981; Ren, 2004) and low (e.g., Fedele et al., 2015; Mahood and Stimac, 1990; Ren, 2004;) values of $D_{Ba}$ and $D_{Sr}$ in different natural rocks (Icenhower and London, 1996; Mahood and Stimac, 1990). Calcium is moderately incompatible in alkali feldspar, with, $D_{Ca}$ values between ~0.60 and 0.95 in short-duration experiments, and lower values in longer experiments (~0.25<$D_{Ca}$<0.45), hence Ca-partitioning is only slightly affected by equilibration time (Figs 5d and 7c, d). Experimental run times do not substantially affect the partitioning of K and Na (Table 6). Our alkali feldspar crystals have stoichiometric compositions within error (Fig. 7a), suggesting that excess Ba and Sr are balanced via Eq. 4 rather than defect substitution. This explains why Onuma diagrams for divalent cations, obtained from experiments with different times, describe always a parabolic trend perfectly fitted by the lattice strain model (Blundy and Wood, 1994).

Time is a fundamental variable affecting crystallization kinetics of magmas. Growth rates are fast at the beginning of the crystallization process and then decrease with increasing time (Arzilli and Carroll, 2013; Iezzi et al., 2014; Lasaga, 1998; Vetere et al., 2013, 2015). Our experimental results show that in particular Ba and Sr partition coefficients depend on growth rate ($Y_L$; Fig. 8), higher growth rates corresponding to higher apparent $D_{Ba}$ and $D_{Sr}$ values (Fig. 8). This result is contrary to experimental data obtained by Long (1978), who observed $D_{Sr}$ decreasing with increasing growth rates, whereas $D_{Ba}$ shows a tenuous inverse relationship with the growth rate of alkali feldspar. These differences can be explained by (i) the increased levels of doping in Long (1978), i.e. our tens of ppm vs. several wt.%, which may not necessarily follow Henry's law, and (ii) by the scatter affecting Long's measurements (see Figs. 2-4 in Long 1978).

In the following, we reason why for alkali-feldspar, departure from equilibrium conditions can significantly affect trace element partitioning during crystal growth (Baker, 2008; Gagnevin et al., 2005; Morgan and London, 2003; Watson and Muller, 2009). Rapid crystal growth of plagioclase and alkali feldspar is often controlled by diffusion at conditions far from the equilibrium (e.g., Dowty, 1980; Hammer, 2008; Kirkpatrick, 1981; Lofgren, 1980), producing hopper and elongated crystal shapes. Disequilibrium growth promotes supersaturation of trace elements in the diffusive boundary layer in the surrounding melt (Albarede and Bottinga, 1972), in which incompatible elements will be enriched, whereas compatible elements will be depleted. Both of these phenomena may be affected by element diffusivities, such that slow moving elements (e.g., Ba) will be more enriched in boundary layer compared with fast moving elements (e.g., Na and K). Because the boundary layer changes composition with time (Watson and Muller, 2009), the value of the specific partition coefficient for element $i$ for the shorter experiments will be different from those of the longer runs.

Therefore, during short-duration experiments, highly compatible trace elements (e.g., Ba and Sr) can be incorporated in alkali feldspar above the equilibrium amount (in agreement with Morgan and London, 2003), resulting in a decreased concentration in the boundary layer (Figs 5a, b and 7c, d). Under disequilibrium conditions, high growth rates of alkali feldspar can favour less selectivity for trace element incorporation, filling up the lattice site as fast as possible and favouring disordered crystal growth (Carpenter and Putnis, 1985) (Fig. 8b). Therefore, highly compatible trace elements, such as Ba and Sr, are more strongly partitioned into the crystal structure than expected for equilibrium conditions (Figs 5a, b and 7d). The substitution mechanism driven by disordered growth, by which trace elements are incorporated into the crystal structure, may control partition coefficients (Corgne and Wood, 2005; Prowatke and Klemme, 2006), promoting high $D_{Ba}$ and $D_{Sr}$ in alkali feldspar under disequilibrium conditions.

As equilibrium is approached, diffusive re-equilibration of cations can occur between alkali feldspar crystals and trachytic melts, promoted by ordered growth (Fig. 8b), which facilitates the selection of compatible elements in the crystal lattice on the basis of the cation charge and size (fitting in the octahedral site). Therefore, the concentrations of highly compatible trace elements (Ba and Sr) will be comparatively low, resulting in lower $D_{Ba}$ and $D_{Sr}$ (Figs 5a, b and 7d) than obtained under disequilibrium conditions.

Most of the trivalent cations are incompatible (D<0.1), their partition coefficients showing maximum values at cation sizes close to La (Fabbrizio et al., 2009). Thus, alkali feldspar is slightly enriched in LREE compared to HREE, in agreement with previous studies (Fabbrizio et al., 2009; Mahood and Hildreth, 1983; Mahood and Stimac, 1990; Ren, 2004; Stix and Gorton, 1990; White et al., 2003). Trivalent cations are also slightly influenced by diffusive re-equilibration time and cation order-disorder growth, as they are slightly enriched in the short-duration experiments, showing higher $D_i$ than those obtained at long-duration experiments (Table 6). Higher values for $D_{Eu}$ probably reflect a mixture of $Eu^{2+}$ and $Eu^{3+}$ in our samples.

*4.6 Implications for natural systems*

Studies of trace element partitioning between alkali feldspar and natural trachytic, phonolitic, pantelleritic and rhyolitic melts have yielded a wide range of values for the distribution coefficients of Rb, Ba, and Sr (Fedele et al., 2015; Pappalardo et al., 2008; Ren, 2004; Villemant 1988; White et al., 2003). The conditions that control these variations have been poorly constrained. Results of this study improve our understanding about timescales and mechanisms that control trace elements mineral/liquid partitioning for highly evolved melts, especially in non-equilibrium conditions. Our study provides new data on the effect of diffusive re-equilibration time and crystal growth rate on trace element partitioning, which allow us to better understand the crystallization kinetics and geochemical evolution of magmas in volcanic systems during pre- and syn-eruptive conditions. Yet, our study cannot provide definitive constraints of crystallization timescales, as other factors such as undercooling and cooling and decompression rates may play an important role in controlling apparent partition coefficients. Nevertheless, here we highlight the effect of the re-equilibration time on element partitioning for static conditions that could represent magma stagnation in the chamber or sill at low degrees of undercooling (between ~30 and 50 °C). Particularly, results of this study have implications on the Campi Flegrei volcanic system (Napoli, Italy), which is an active volcanic caldera characterized by evolved alkaline rocks (trachytes, trachyphonolites and phonolites) with alkali feldspar as the dominant crystal phase (e.g., Arzilli et al., 2016; Fedele et al. 2015; Piochi et al. 2005). The compositions of alkali feldspar crystals in trachyphonolitic and phonolitic magmas of Campi Flegrei (Fedele et al., 2015; Pappalardo et al., 2008; Villemant, 1988) are similar to those obtained from this experimental work, with Or contents of 32-57 mol.%. The partition coefficients of Ba between alkali feldspar and trachyphonolite melts of Campi Flegrei show large variations (Fedele et al., 2015; Pappalardo et al., 2008; Villemant, 1988). The partition coefficients of Ba obtained from Villemant (1988) and Pappalardo et al. (2008) range between ~7 and 10, which are comparable to our equilibrium $D_{Ba}$. Instead, data reported by Fedele et al. (2015) show $D_{Ba} \geq 20$, comparable to those obtained from short-duration experiments. These high partition coefficients for Ba may be related to kinetic effects (e.g., diffusive re-equilibration time and crystal growth rates), as disequilibrium crystallization of alkali feldspar is expected to occur during Campi Flegrei eruptions (Arzilli et al. 2016). Crystallization timescales of alkali feldspar in natural trachytic melts can be estimated using the curve fits of $D_{Ba}$ and $D_{Sr}$ with time (Fig. 7d), through the following equations:

$$t = \frac{56.44(\pm 7.68)\ D_{Ba}}{0.228(\pm 0.072)} \quad \text{Eq. 5}$$

$$t = \frac{26.86(\pm 2.84)\ D_{Sr}}{0.102(\pm 0.027)} \quad \text{Eq. 6}$$

where $t$ is time in hours. High $D_{Ba}$ and $D_{Sr}$ (~20 and ~9 respectively) obtained from Fedele et al. (2015) can be produced in ~6 days (Table 10), which may represent the maximum residence time of Campi Flegrei's magma under disequilibrium conditions. In contrast, phenocrysts of alkali feldspar characterized by low $D_{Ba}$ and $D_{Sr}$ (<10 and <3 respectively) suggest residence time ≥9 days (Table 10), highlighting that trachytic systems may reach equilibrium conditions after ~9 days. This suggests that evidences of disequilibrium processes in Campi Flegrei rocks could indicate that the timescale at subliquidus conditions in a reservoir is shorter than 6 days (in agreement with Iezzi et al., 2008). However, further partitioning experiments, investigating the effect of undercooling and decompression rate, should be performed to obtain a complete picture of trace element partitioning variation in natural volcanic samples.

Our results also highlight that the transition from disequilibrium to equilibrium conditions in trachytic and basaltic magmatic systems occurs over different timescales. In fact, plagioclase crystallization in basaltic melts reaches equilibrium in a few hours (Agostini et al., 2013; Arzilli et al., 2015; Kolzenburg et al., 2016; La Spina et al., 2016; Vetere et al., 2015), whereas, alkali feldspar in trachytes in several days, a result of differences in temperature and melt polymerization or diffusivity.

## 5. Concluding remarks

High-silica and alkali-rich magmas are often characterized by explosive eruption styles, therefore the determination of kinetic effects on mineral/liquid partition coefficients for highly evolved systems can significantly improve our understanding of magma evolution processes. This should allow the development of new magma chamber and conduit models that take into account disequilibrium crystallization and magma evolution in the conduit and hence constrain timescales of magmatic processes, facilitating the evaluation of volcanic hazards.

The present study provides new data on partition coefficients for alkali feldspar, showing that kinetic effects involving diffusive re-equilibration, growth rate and order-disorder crystal growth influence element partitioning between alkali-feldspar and trachy-phonolitic melts. The rapid growth of alkali feldspar under disequilibrium conditions contributes to less selectivity in partitioning of compatible trace elements (e.g., Ba and Sr), as disordered crystal growth is the favourite process during the crystallization. This produces higher apparent $D_{Ba}$ and $D_{Sr}$ than those obtained at equilibrium conditions, which then allow for identifying disequilibrium crystallization and which should ultimately constrain time scales of magmatic processes.

**Acknowledgements**

This work was supported by PRIN 2009 (2009PZ47NA_002), FAR2012 (M. R. Carroll) and by ERC Consolidator Grant 612776-CHRONOS. A. Fabbrizio acknowledges support from the Czech Science Foundation (GACR, project 18-01982S). We also wish to thank P. Ulmer and E. Reusser (ETH, Zurich) and Raul Carampin (CNR-IGG, Padova) for precious advice during microprobe analysis. We are grateful to U. Mann for assistance during experiments, and M. Pistone, L. Martin (ETH, Zurich) and J.M. Henot (LMV, Blaise Pascal University) for helpful advices during SEM analysis. We thank G. Iezzi, M. Ren and two anonymous reviewers for constructive and helpful comments. We would like to thank J. Blundy for stimulating discussions and helpful suggestions.

**Figure captions**

Figure 1. Alkali feldspar liquidus diagram for ZAC composition at water-saturated and -undersaturated conditions. Reported experimental data are from Fabbrizio and Carroll (2008) and Arzilli and Carroll (2013). The liquidus of alkali feldspar, related to water-undersaturated conditions (2 wt. % of $H_2O$ dissolved in the melt), has a positive slope with a gradient of 10 °C/100 MPa (the gradient is extrapolated from data of Burnham, 1979). Closed circles: alkali feldspar (af) + clinopyroxene (cpx) + biotite (bt) + magnetite (mt); closed diamonds: alkali feldspar (af) + clinopyroxene (cpx) + magnetite (mt); half-open squares: clinopyroxene (cpx) + biotite (bt) + magnetite (mt); open squares: clinopyroxene (cpx) + magnetite (mt); open diamonds: magnetite (mt).

Figure 2. BSE images of short and long duration experiments. (a) Texture of sample T2 (24 hours) shows alkali feldspar (af) crystals (darker phase) with tabular, prismatic elongated, acicular and hopper shapes hosted by glassy matrix (gl). (b) Enlarged view of sample T2 showing tabular and hopper morphologies of alkali feldspar promoted by rapid growth. (c) Texture of sample T11 (198 hours) shows prismatic crystals of alkali feldspar (af) together with minor amount of clinopyroxene (cpx). (d) Enlarged view of sample T11 showing alkali feldspar (af), clinopyroxene (cpx) and glass

(gl).

Figure 3. Onuma diagrams for alkali-feldspar grown during short-duration experiments. The partition coefficients were calculated for mono-, di-, tri-, and quadrivalent cations entering the X-fold alkali-feldspar A-site. Parabolas are calculated from the partition coefficients for monovalent (circles) and selected divalent (squares) cations. Ionic radii (X-fold coordination) are taken from Shannon (1976). Onuma diagrams are shown for sample: (a) T8 ($Or_{46.4}$), (b) T6 ($Or_{49.9}$), (c) T2 ($Or_{50.2}$), (d) T9 ($Or_{55.9}$) and (e) T10 ($Or_{57.0}$).

Figure 4. Onuma diagrams for alkali-feldspar grown during long-duration experiments. The partition coefficients were calculated for mono-, di-, tri-, and quadrivalent cations entering the X-fold alkali-feldspar A-site. Parabolas are calculated from the partition coefficients for monovalent (circles) and selected divalent (squares) cations. Ionic radii (X-fold coordination) are taken from Shannon (1976). Onuma diagrams are shown for sample: (a) T11 ($Or_{32.3}$) and (b) T12 ($Or_{47.0}$).

Figure 5. Correlation between partition coefficients of Ba, Sr, Rb and Ca as a function of Or content of alkali feldspar. (a) $D_{Ba}$ vs Or (squares). (b) $D_{Sr}$ vs Or (circles). (c) $D_{Rb}$ vs Or (diamonds). (d) $D_{Ca}$ vs Or (triangles). The diagram shows the comparison between partition coefficients obtained through short- and long-duration experiments. Open symbols: partition coefficients values obtained from long-duration experiments; closed symbols: partition coefficients values obtained from short-duration experiments.

Figure 6. (a) Comparison of the alkali-feldspar/silicate liquid distribution coefficients, as function

of atomic number, for synthetic systems. Open diamonds: Icenhower and London (1996); open triangles: Morgan and London (2003); open reversed triangles: Henderson and Pierozynski (2012); open squares: Fabbrizio et al., (2008); closed circles: this study. Melt compositions used by different studies: peraluminous granite (Icenhower and London, 1996); granite (Morgan and London, 2003); trachyte and phonolite (Henderson and Pierozynski, 2012); rhyolite (Fabbrizio et al., 2008). (b) Comparison between the alkali-feldspar/silicate liquid distribution coefficients, as function of atomic number, from this study and from trachytic natural rocks. The symbols indicate the minimum and the maximum value for the analyzed element; the vertical solid line indicates the variation range for the specific element. Open squares: selected data for natural trachytic compositions from Pappalardo et al., (1988); Villemant (1998); White et al., (2003); Fedele et al., (2015). Closed circles: data from this study.

Figure 7. (a) Plot of $M^+$ (sum of K, Na, Rb and Cs) + $Si^{4+}$ vs. $M^{2+}$ (sum of Ca, Ba, and Sr) + $Al^{3+}$ (cations) in the synthesized alkali-feldspar. The solid line represents the ideal substitution $M^{1+} + Si^{4+}$ = - ($M^{2+}$ + $Al^{3+}$) + 5. (b) Plot of Ba, Sr and Ca content (ppm) in alkali feldspar as a function of the orthoclase contents in mol.%. (c) Plot of Ba, Sr and Ca content (ppm) in alkali feldspar as a function of the experimental time (hours). (d) Semilogarithmic plot of the partition coefficient values for Ba (squares), Sr (circles), and Ca (triangles) as a function of the experimental time (hours). Note the decrease in the partition coefficient values with increasing time.

Figure 8. (a) Logarithmic plot of the partition coefficient values of Ba (closed squares), Sr (open circles), Ca (closed triangles) and Rb (open squares) as a function of alkali feldspar growth rate. (b) The relationship between the strain-compensated partition coefficients ($D_0$) and Or content shows

that disordered growth is favoured at short-duration experiments. Low strain-compensated partition coefficients imply that ordered growth mechanism is facilitated in long-duration experiments.

# Equations

$$\log D_i = \log D_0 + 0.43429 \times \left\{ \frac{-91.017 \times E \times \left[ \frac{r_0}{2} \times (r_i - r_0)^2 + \frac{1}{3}(r_i - r_0)^3 \right]}{T} \right\} \qquad \text{Eq. 1}$$

$$r_c^{2+} = 1.341 + 0.207 X_{Or} \ (\text{Å}) \qquad \text{Eq. 2}$$

$$E_c = 1125 \times valence \times \left( r_c^{n+} + 1.38 \right)^{-3} \ (\text{GPa}) \qquad \text{Eq. 3}$$

$$M^{1+} + Si^{4+} = M^{2+} + Al^{3+} \qquad \text{Eq. 4}$$

$$t = \frac{56.44(\pm 7.68) - D_{Ba}}{0.228(\pm 0.072)} \qquad \text{Eq. 5}$$

$$t = \frac{26.86(\pm 2.84) - D_{Sr}}{0.102(\pm 0.027)} \qquad \text{Eq. 6}$$

**Figure 1**
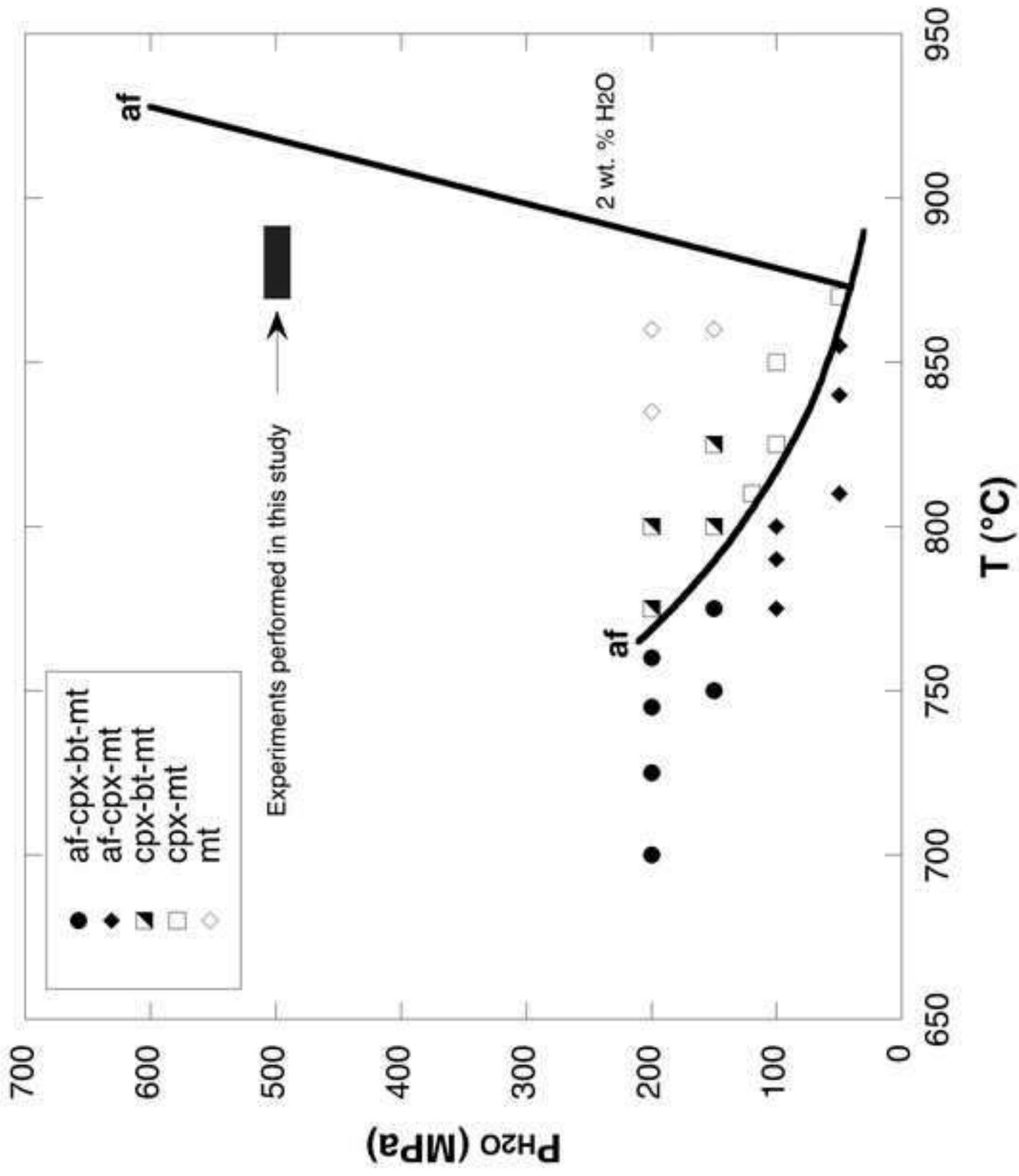

**Figure 2**
[Click here to download high resolution image](#)

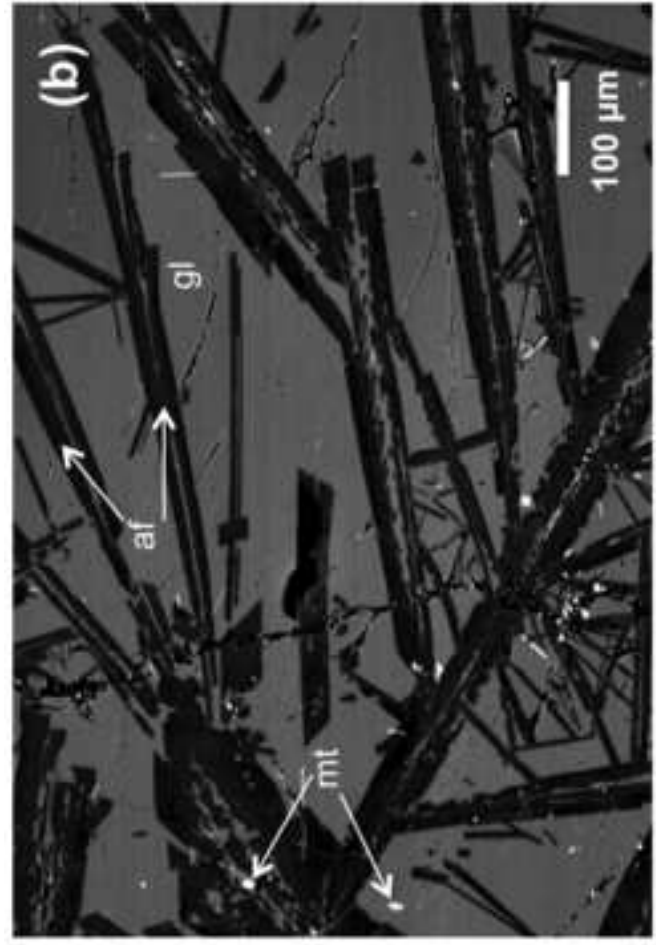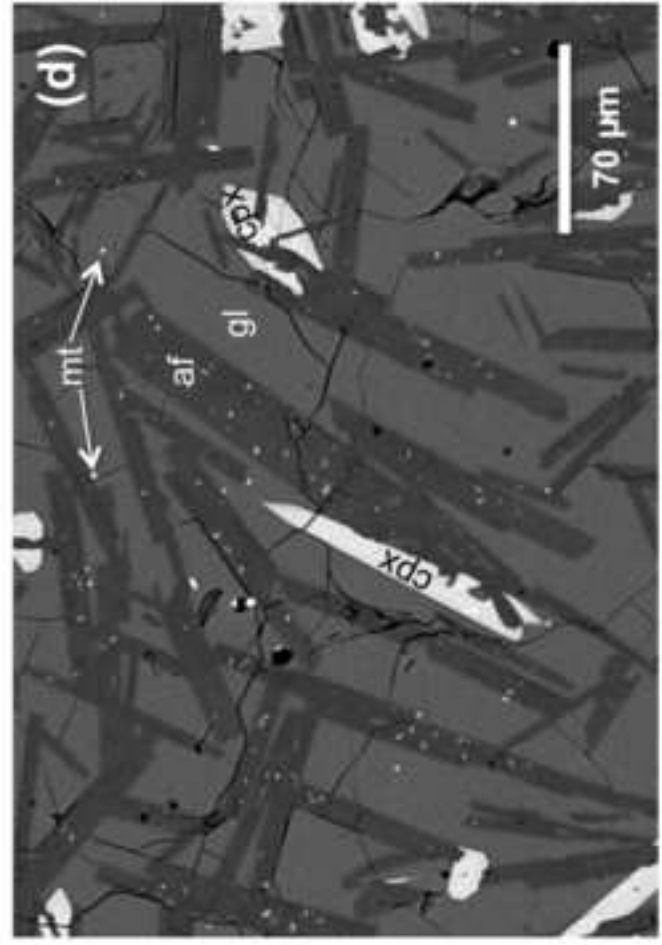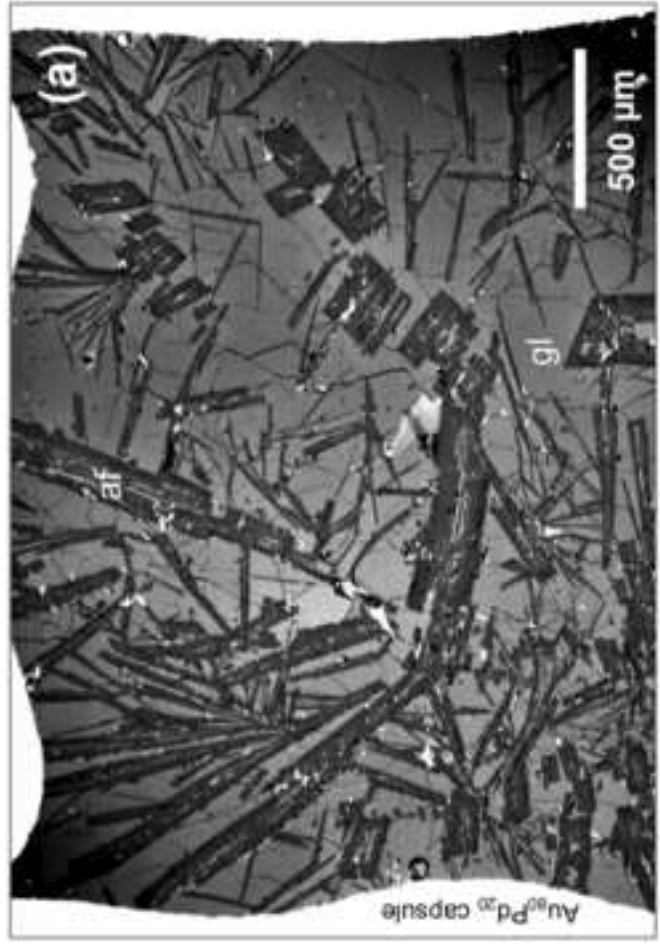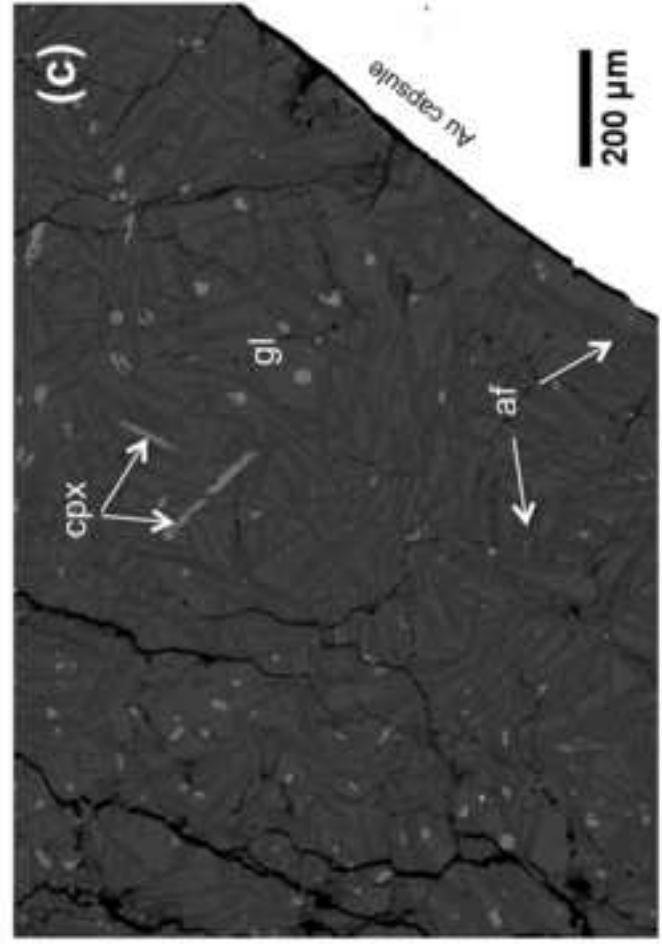

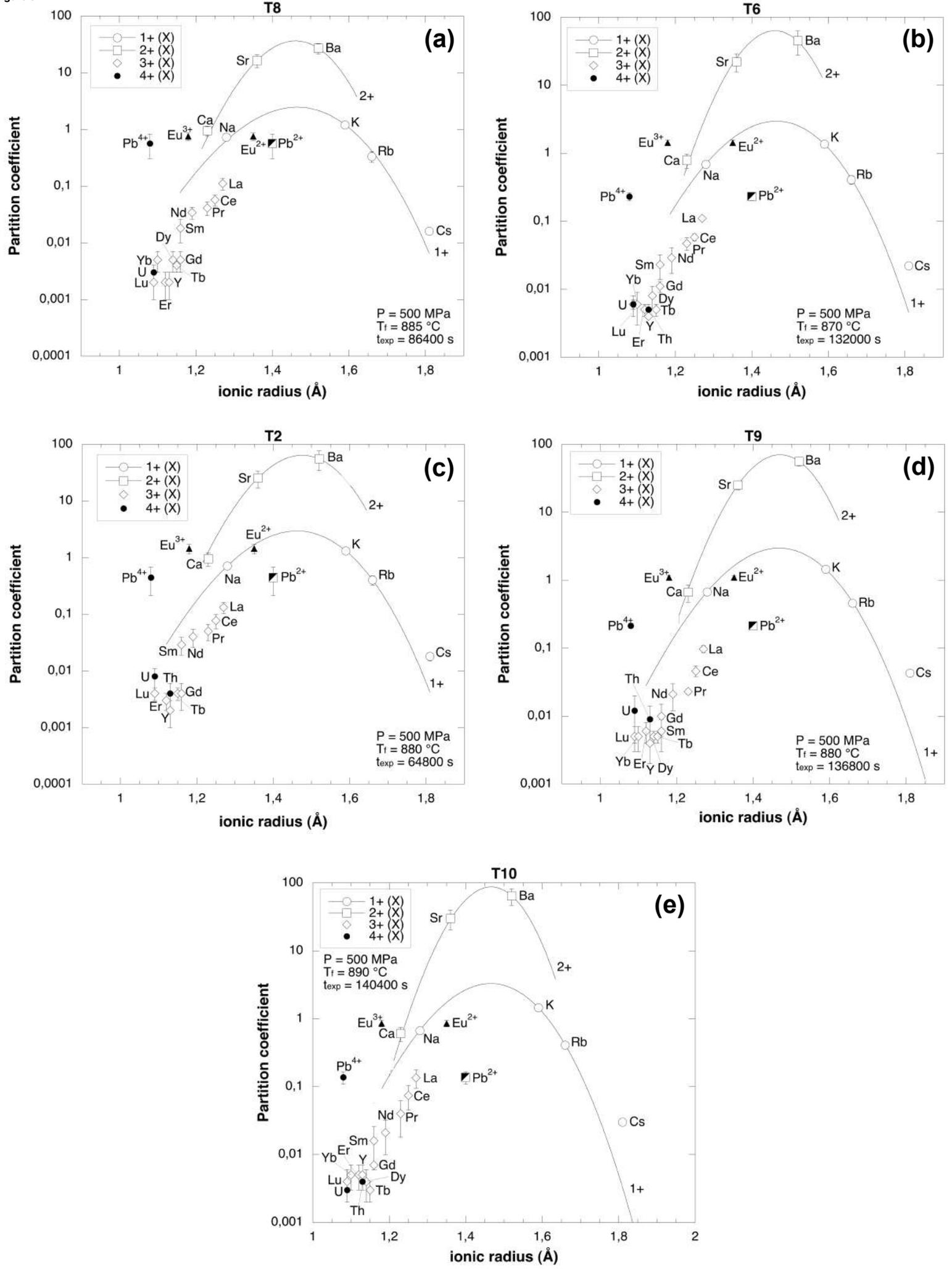

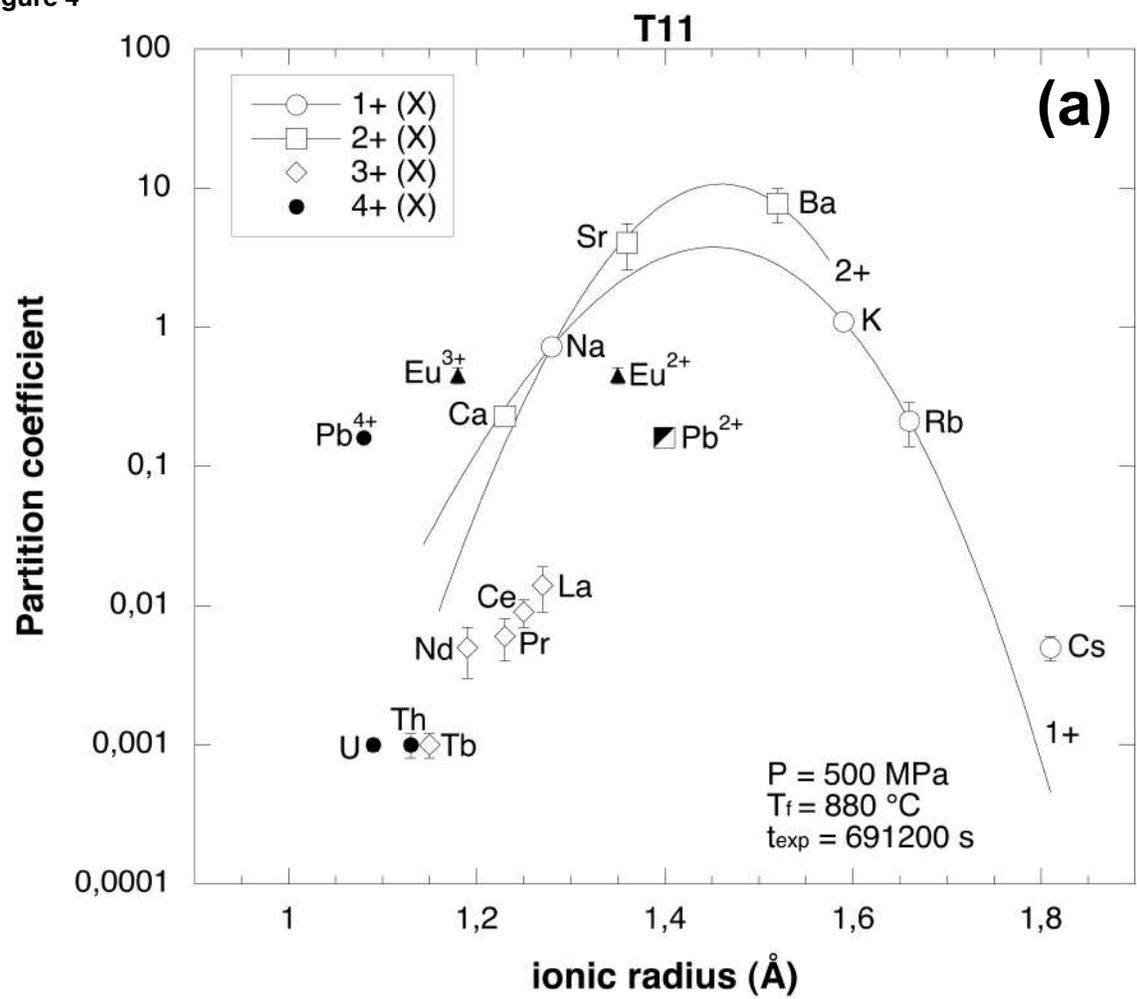
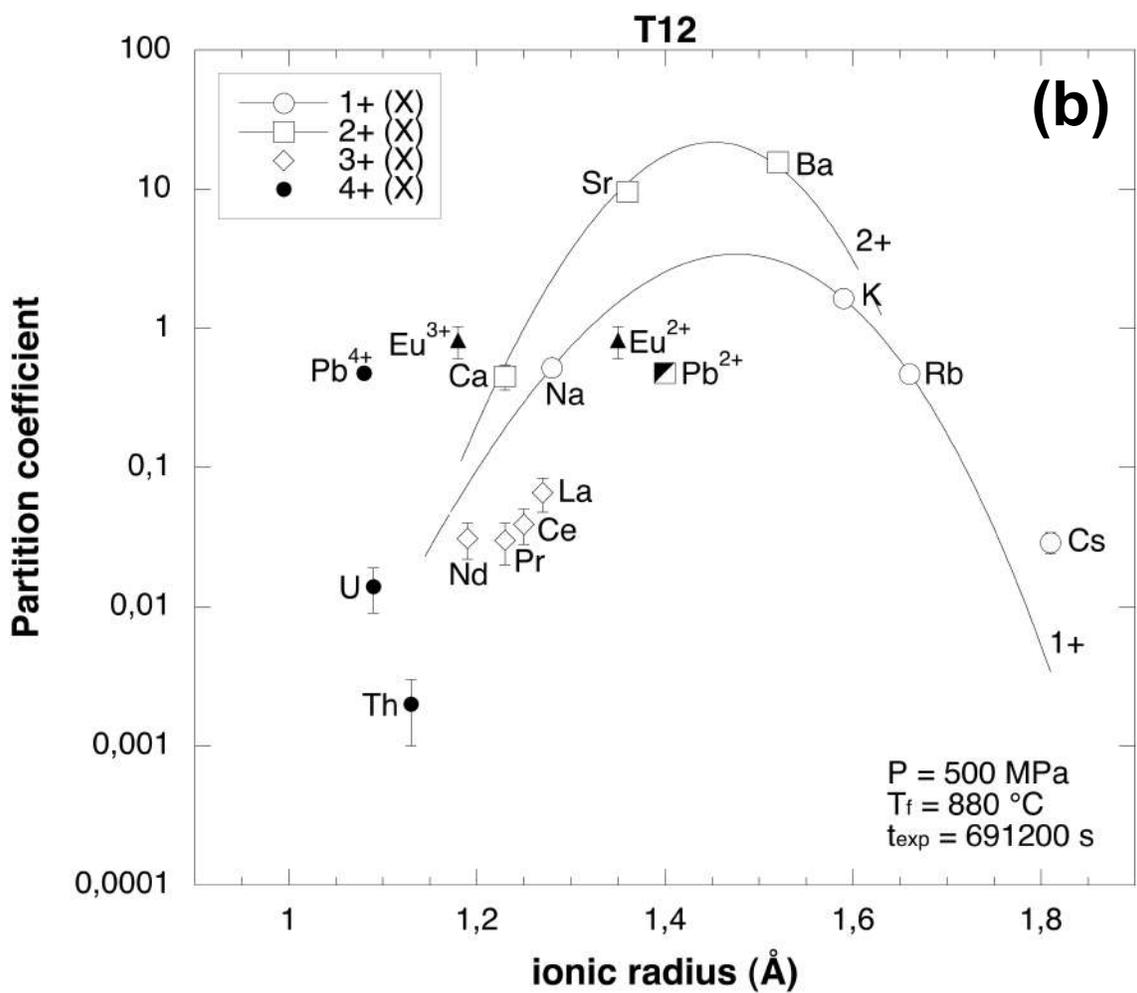

Figure 5

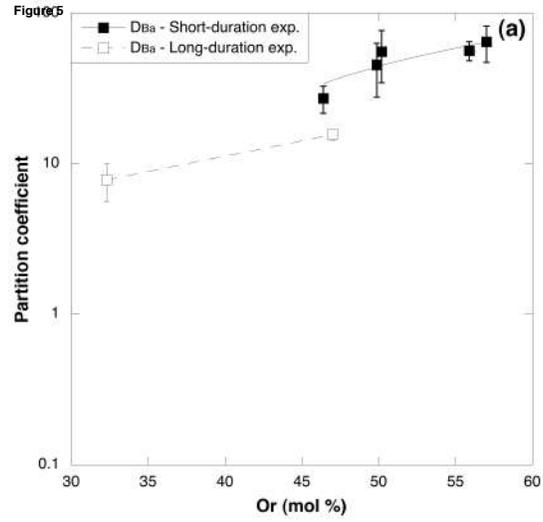

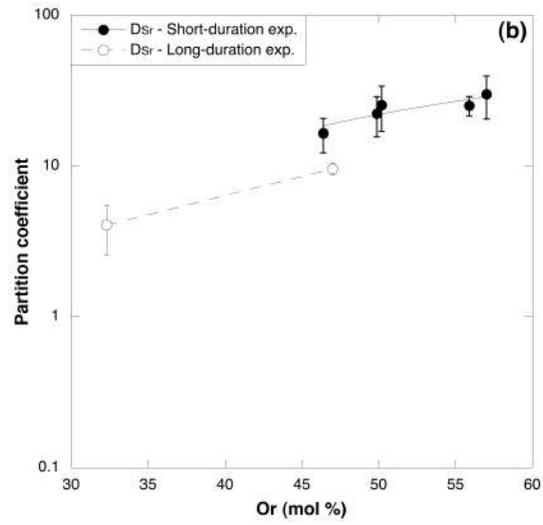

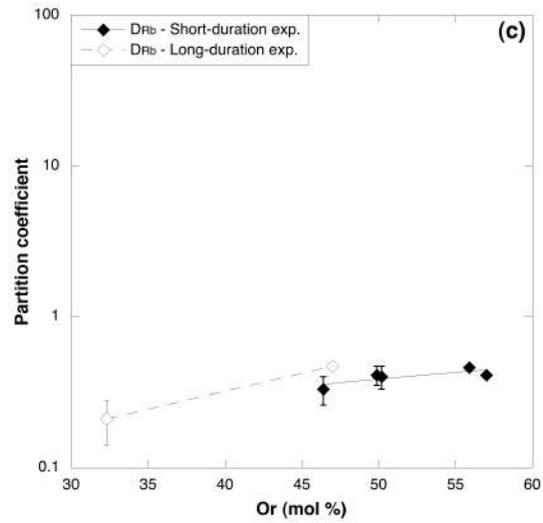

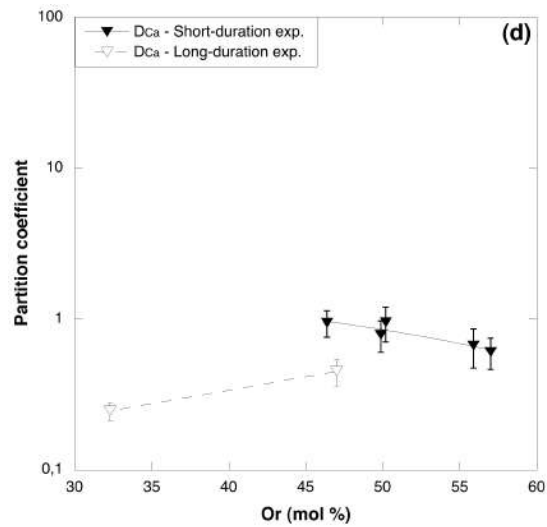

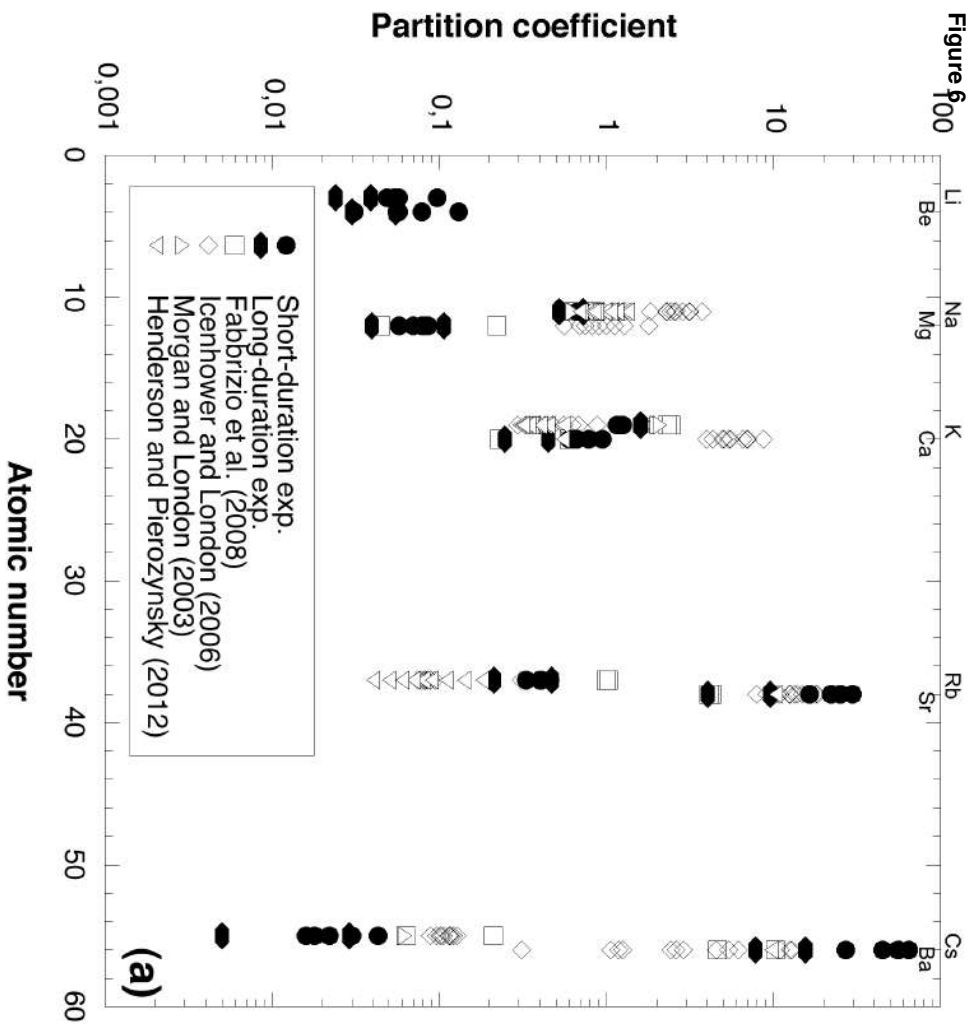
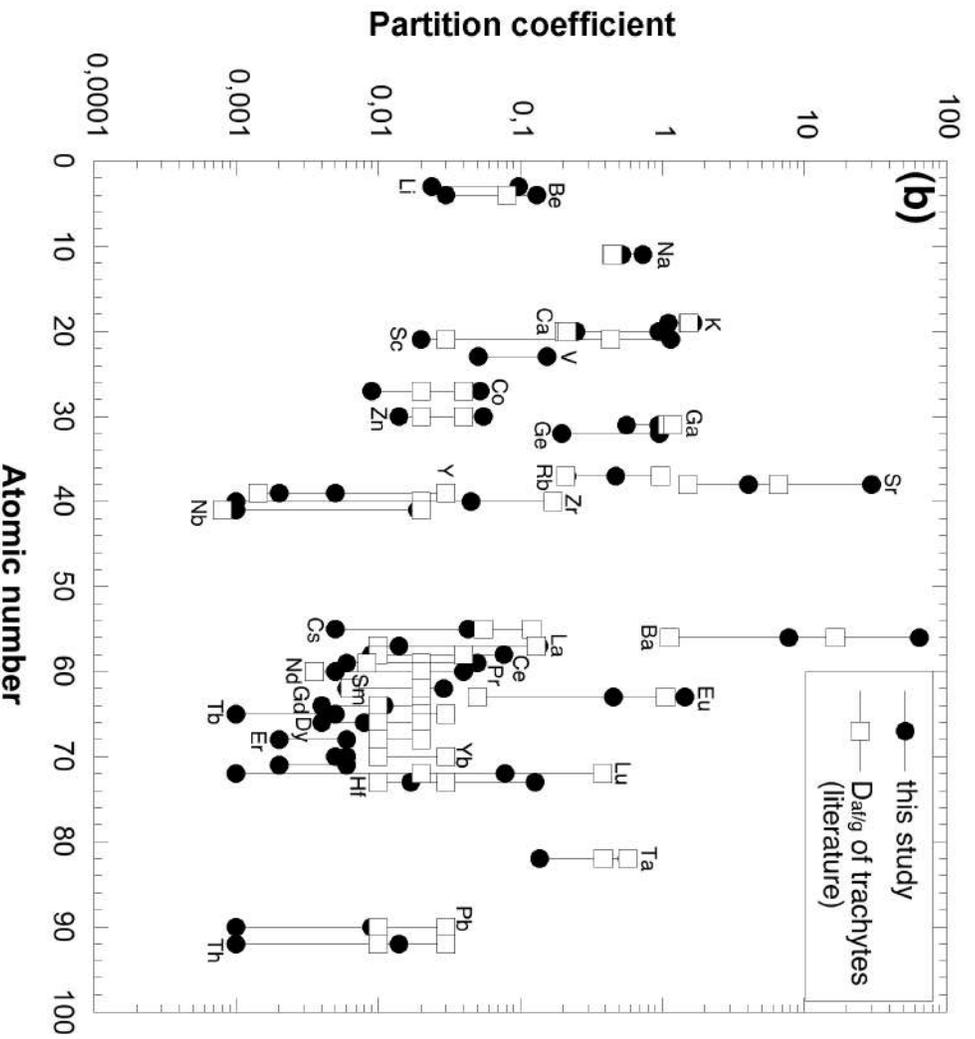

Figure 6

Figure 7

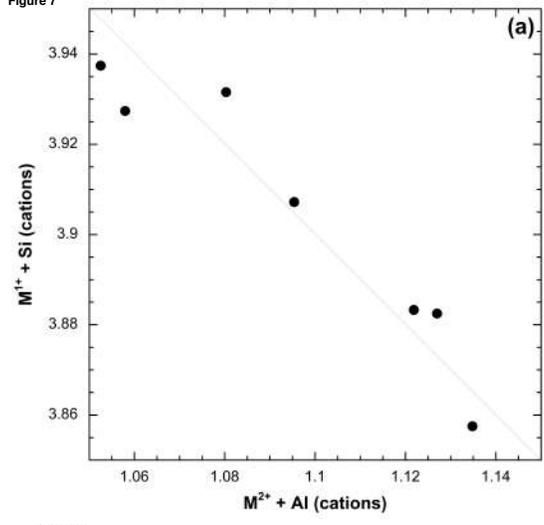
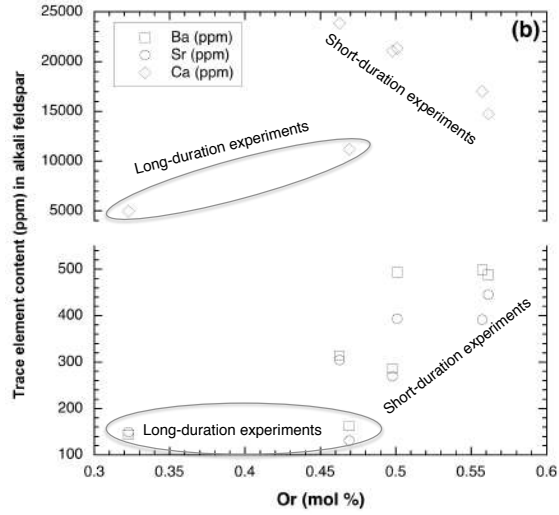
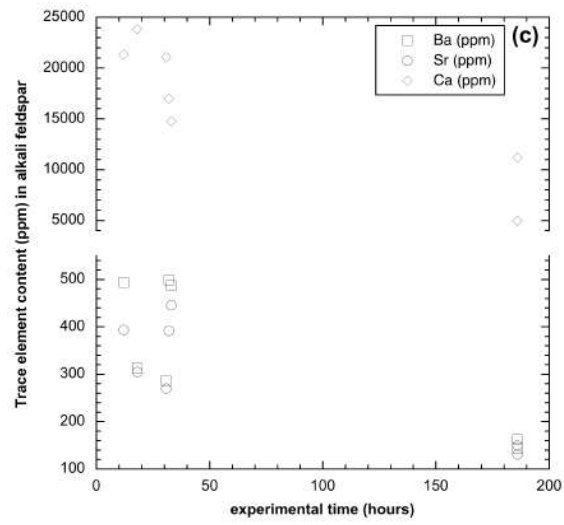
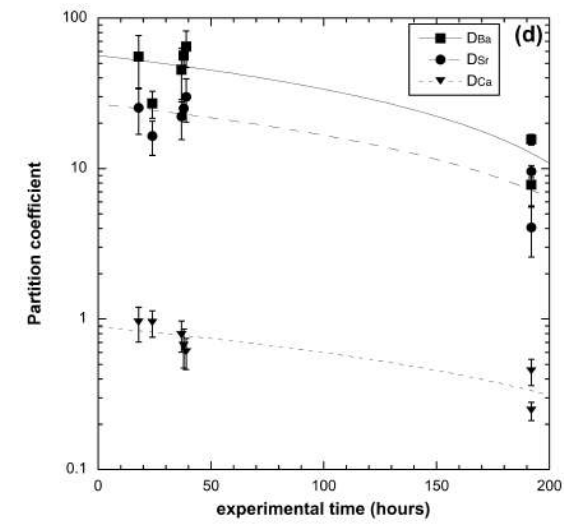

Figure 8

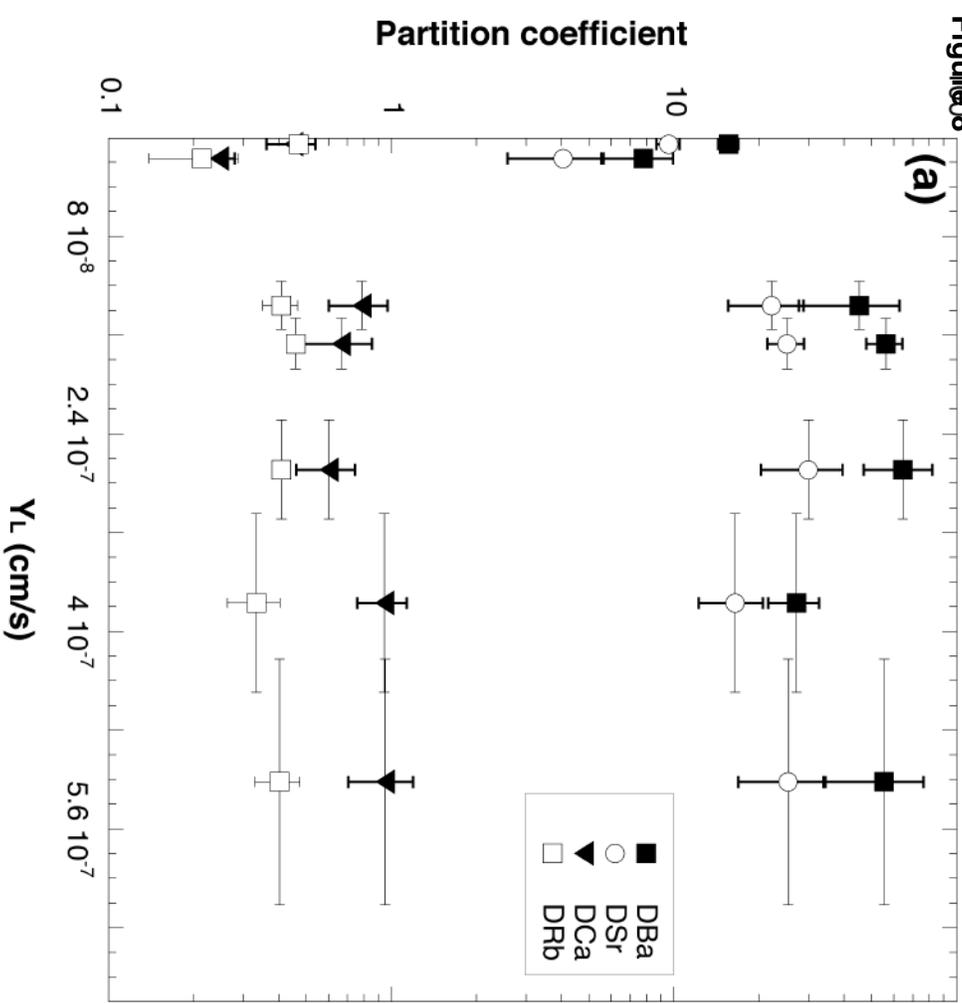

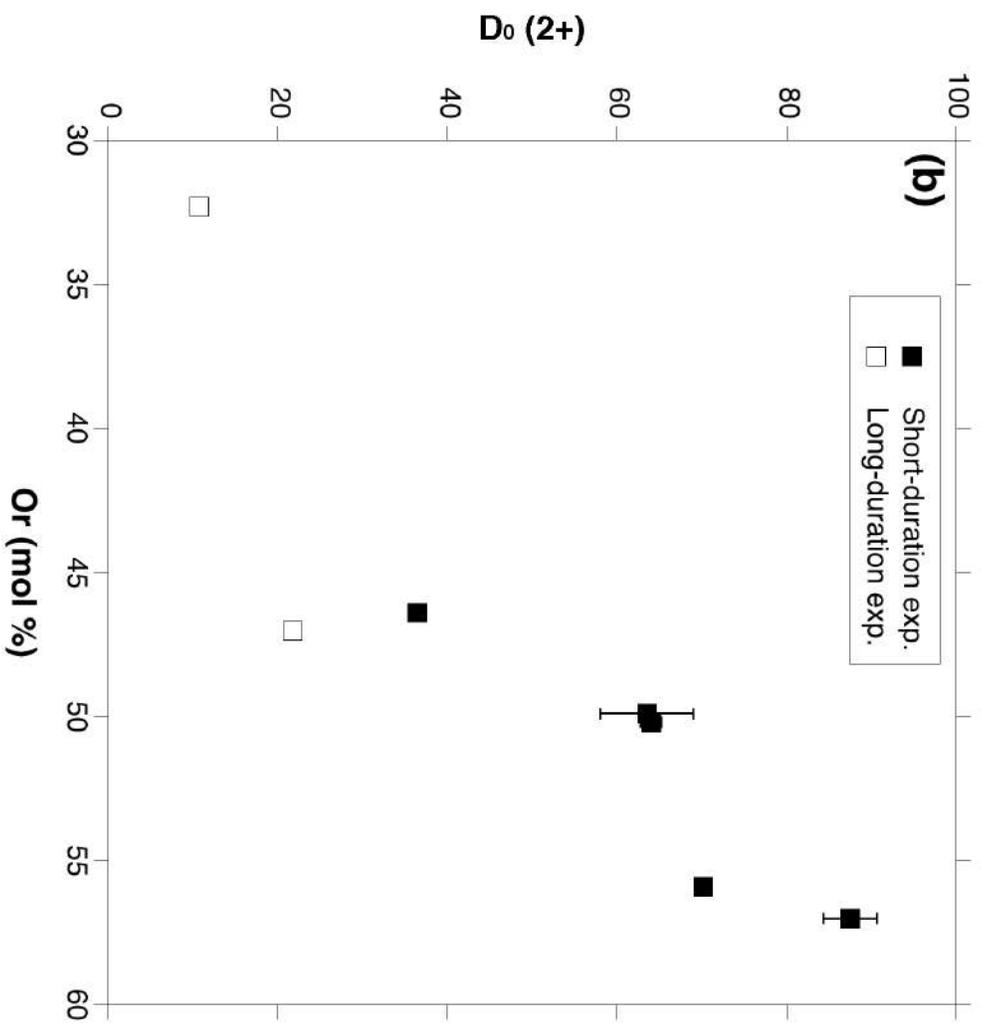

Table 1: Chemical composition of ZAC (from Di Matteo et al. 2004).

| oxide (wt%) | ZAC |
|---|---|
| $SiO_2$ | 62.18 |
| $TiO_2$ | 0.45 |
| $Al_2O_3$ | 18.70 |
| FeO* | 3.19 |
| MnO | 0.27 |
| MgO | 0.23 |
| CaO | 1.65 |
| $Na_2O$ | 6.16 |
| $K_2O$ | 7.14 |
| $P_2O_5$ | 0.02 |
| Total | 100.00 |

note: FeO* = total iron as FeO. Analysis performed by X-ray fluorescence (XRF) and normalized to 100%; original analytical total = 99.25.

Table 2. Experimental conditions of short and long duration experiments.

| sample | P (MPa) | *$H_2O$ (wt %) | $T_i$ (°C) | $T_f$ (°C) | $t_m$ (s) | $t_{exp}$ (s) |
|---|---|---|---|---|---|---|
| *Short duration experiments* | | | | | | |
| T1 | 500 | 2 | 1100 | 1100 | 21600 | 0 |
| T2 | 500 | 2 | 1100 | 880 | 21600 | 64800 |
| T8 | 500 | 2 | 1100 | 885 | 21600 | 86400 |
| T6 | 500 | 2 | 1100 | 870 | 21600 | 132000 |
| T9 | 500 | 2 | 1100 | 880 | 21600 | 136800 |
| T10 | 500 | 2 | 1100 | 890 | 21600 | 140400 |
| *Long duration experiments* | | | | | | |
| T11 | 500 | 2 | 1040 | 880 | 21600 | 691200 |
| T12 | 500 | 2 | 1040 | 880 | 21600 | 691200 |

note: P = experimental pressure; * $H_2O$ = water content dissolved in the melt; $T_i$ = initial (or melting) temperature; $T_f$ = final (or experimental) temperature; $t_m$ = melting time; $t_{exp}$ = experimental time.

Table 3. Growth rates of alkali feldspar obtained through short and long duration experiments.

| Sample | $Y_L$ (cm/s) | $\sigma Y_L$ |
|---|---|---|
| *Short duration experiments* | | |
| T2 | 5.22E-07 | 9.96E-08 |
| T8 | 3.77E-07 | 7.24E-08 |
| T6 | 1.36E-07 | 1.95E-08 |
| T9 | 1.67E-07 | 2.07E-08 |
| T10 | 2.69E-07 | 4.02E-08 |
| *Long duration experiments* | | |
| T11 | 1.68E-08 | 2.45E-09 |
| T12 | 5.42E-09 | 1.23E-09 |

note: $Y_L$: growth rate of alkali feldspar. s $Y_L$: standard deviation of growth rate of alkali feldspar.

Table 4. Average chemical compositions of alkali feldspars. Major and trace elements were measured using Electron Microprobe and Laser Ablation ICP-MS respectively.

|  | *Short duration experiments* | | | | | | | | | | *Long duration experiments* | | | |
|---|---|---|---|---|---|---|---|---|---|---|---|---|---|---|
|  | T2 | σ | T8 | σ | T6 | σ | T9 | σ | T10 | σ | T11 | σ | T12 | σ |
| $SiO_2$ (wt %) | 64.09 | 0.47 | 63.85 | 0.36 | 63.93 | 0.42 | 64.54 | 0.60 | 64.34 | 0.39 | 66.29 | 0.90 | 66.36 | 0.59 |
| $TiO_2$ (wt %) | 0.15 | 0.03 | 0.15 | 0.04 | 0.17 | 0.04 | 0.15 | 0.05 | 0.14 | 0.04 | 0.06 | 0.01 | 0.08 | 0.03 |
| $Al_2O_3$ (wt %) | 19.58 | 0.42 | 19.51 | 0.33 | 19.44 | 0.30 | 19.28 | 0.24 | 18.79 | 0.25 | 19.63 | 0.49 | 19.35 | 0.44 |
| FeO (wt %) | 0.28 | 0.06 | 0.47 | 0.09 | 0.33 | 0.08 | 0.28 | 0.05 | 0.40 | 0.08 | 0.20 | 0.04 | 0.30 | 0.13 |
| MnO (wt %) | 0.02 | 0.01 | 0.02 | 0.01 | 0.02 | 0.01 | 0.03 | 0.02 | 0.02 | 0.01 | <0.01 | n.a. | <0.01 | n.a. |
| MgO (wt %) | 0.02 | 0.01 | 0.03 | 0.02 | 0.02 | 0.01 | 0.02 | 0.01 | 0.02 | 0.01 | <0.01 | n.a. | <0.03 | n.a. |
| CaO (wt %) | 1.49 | 0.38 | 1.67 | 0.31 | 1.48 | 0.34 | 1.19 | 0.33 | 1.03 | 0.23 | 0.35 | 0.05 | 0.79 | 0.15 |
| $Na_2O$ (wt %) | 4.99 | 0.23 | 5.24 | 0.38 | 5.01 | 0.32 | 4.46 | 0.17 | 4.44 | 0.31 | 7.52 | 0.42 | 5.67 | 0.55 |
| $K_2O$ (wt %) | 8.92 | 0.57 | 8.10 | 0.82 | 8.81 | 0.70 | 9.85 | 0.46 | 10.10 | 0.61 | 5.59 | 0.62 | 8.22 | 0.96 |
| Total | 99.55 | 0.25 | 99.03 | 0.27 | 99.21 | 0.31 | 99.79 | 0.46 | 99.28 | 0.32 | 99.64 |  | 100.77 |  |
| An | 7.0 |  | 8.0 |  | 7.0 |  | 5.7 |  | 4.9 |  | 1.7 |  | 3.8 |  |
| Ab | 42.8 |  | 45.6 |  | 43.1 |  | 38.4 |  | 38.1 |  | 66.0 |  | 49.3 |  |
| Or | 50.2 |  | 46.4 |  | 49.9 |  | 55.9 |  | 57.0 |  | 32.3 |  | 47.0 |  |
| Na (ppm) | 37043 | 1709 | 38841 | 2833 | 37162 | 2386 | 33060 | 1282 | 32945 | 2312 | 55756 | 3080 | 42099 | 4091 |
| K (ppm) | 74017 | 4761 | 67232 | 6789 | 73142 | 5823 | 81787 | 3791 | 83875 | 5097 | 46442 | 5143 | 68275 | 7930 |
| Ca (ppm) | 21359 | 5415 | 23875 | 4441 | 21089 | 4885 | 17023 | 4675 | 14774 | 3353 | 4985 | 684 | 11227 | 2187 |
| Li (ppm) | 14 | 1 | 12 | 2 | 14 | 4 | 8.6 | 0.3 | 18 | 10 | 1.5 | 0.4 | 4.1 | 0.3 |
| Be (ppm) | 11 | 6 | 14 | 4 | 17 | 4 | 4.0 | 2.1 | 12 | 7 | 15 | 8 | 1.3 | 0.5 |
| Sc (ppm) | 5.6 | 0.3 | 6.3 | 2.8 | 4.3 | 0.8 | 2.0 | 0.1 | 3.3 | 0.6 | 7.4 | 1.4 | 12 | 0 |
| V (ppm) | 11 | 2 | 9.3 | 6.7 | 12 | 4 | 5.3 | 0.1 | 9.1 | 4.3 | 9.2 | 2.5 | <0.50 | n.a. |
| Co (ppm) | 3.3 | 1.2 | 0.73 | 0.21 | 0.69 | 0.25 | 0.55 | 0.18 | 1.1 | 0.3 | 1.1 | 0.6 | <0.24 | n.a. |
| Zn (ppm) | 6.7 | 1.5 | 3.6 | 2.1 | 2.7 | 1.3 | 2.2 | 1.3 | 2.2 | 0.8 | 2.6 | 0.1 | <6.4 | n.a. |
| Ga (ppm) | 70 | 5 | 103 | 9 | 50 | 10 | 63 | 4 | 113 | 1 | 92 | 12 | 14 | 1 |
| Ge (ppm) | 49 | 10 | 53 | 24 | 6.5 | 2.6 | 30 | 0 | 71 | 20 | 64 | 5 | 6.3 | 2.4 |
| Rb (ppm) | 264 | 47 | 208 | 43 | 268 | 36 | 263 | 6 | 228 | 1 | 106 | 37 | 243 | 5 |
| Sr (ppm) | 394 | 79 | 305 | 51 | 270 | 43 | 392 | 51 | 446 | 34 | 150 | 53 | 131 | 12 |
| Y (ppm) | 0.33 | 0.09 | 0.30 | 0.08 | 0.60 | 0.06 | 0.64 | 0.30 | 0.95 | 0.49 | <0.05 | n.a. | <2.05 | n.a. |
| Zr (ppm) | 6.0 | 1.8 | 7.5 | 0.6 | 49 | 7 | 15 | 2 | 10 | 2 | 1.3 | 0.6 | 34 | 2 |
| Nb (ppm) | 3.0 | 0.0 | 1.9 | 0.3 | 4.8 | 0.4 | 4.3 | 0.9 | 2.9 | 0.3 | 0.72 | 0.07 | 4.2 | 0.6 |
| Cs (ppm) | 8.5 | 0.7 | 5.3 | 0.3 | 7.7 | 0.3 | 11 | 1 | 9.4 | 0.4 | 1.6 | 0.2 | 1.8 | 0.3 |
| Ba (ppm) | 494 | 154 | 314 | 17 | 287 | 62 | 499 | 42 | 488 | 37 | 144 | 27 | 163 | 5 |
| La (ppm) | 35 | 3 | 32 | 5 | 32 | 2 | 23 | 1 | 42 | 13 | 5.3 | 1.9 | 17 | 1 |
| Ce (ppm) | 29 | 6 | 26 | 4 | 29 | 3 | 17 | 2 | 32 | 13 | 5.1 | 0.8 | 19 | 1 |
| Pr (ppm) | 6.4 | 0.9 | 5.3 | 0.6 | 4.9 | 0.9 | 2.3 | 0.0 | 6.2 | 3.4 | 0.94 | 0.22 | 1.5 | 0.2 |
| Nd (ppm) | 7.7 | 2.2 | 7.2 | 0.9 | 6.1 | 2.6 | 3.4 | 1.4 | 4.9 | 2.7 | 1.2 | 0.4 | 5.1 | 0.1 |
| Sm (ppm) | 3.9 | 0.6 | 2.5 | 0.8 | 2.3 | 0.9 | 0.58 | 0.33 | 2.7 | 1.8 | <0.44 | n.a. | <0.68 | n.a. |
| Eu (ppm) | 113 | 21 | 55 | 5 | 55 | 4 | 60 | 3 | 74 | 8 | 42 | 5 | 1.6 | 0.2 |
| Gd (ppm) | 0.58 | 0.31 | 0.59 | 0.27 | 1.05 | 0.22 | 0.85 | 0.40 | 1.14 | 0.03 | <0.2 | n.a. | <0.55 | n.a. |
| Tb (ppm) | 0.56 | 0.17 | 0.50 | 0.09 | 0.34 | 0.07 | 0.39 | 0.08 | 0.37 | 0.08 | 0.14 | 0.01 | <0.05 | n.a. |
| Dy (ppm) | 0.24 | 0.17 | 0.79 | 0.24 | 0.79 | 0.27 | 0.54 | 0.02 | 0.62 | 0.34 | <0.29 | n.a. | <0.48 | n.a. |
| Er (ppm) | 0.47 | 0.05 | 0.25 | 0.15 | 0.42 | 0.09 | 0.46 | 0.14 | 0.75 | 0.32 | <0.176 | n.a. | <0.36 | n.a. |
| Yb (ppm) | <0.35 | n.a. | 0.83 | 0.32 | 0.59 | 0.26 | 0.48 | 0.23 | 0.92 | 0.41 | <0.27 | n.a. | <0.37 | n.a. |
| Lu (ppm) | 0.59 | 0.15 | 0.33 | 0.08 | 0.52 | 0.09 | 0.46 | 0.20 | 0.79 | 0.37 | <0.126 | n.a. | <0.04 | n.a. |
| Hf (ppm) | 2.6 | 0.4 | 8.1 | 3.2 | 8.5 | 1.4 | 1.9 | 0.4 | 2.5 | 0.8 | 0.23 | 0.04 | <0.67 | n.a. |
| Ta (ppm) | 0.72 | 0.26 | 0.40 | 0.15 | 0.49 | 0.04 | 0.11 | 0.03 | 0.20 | 0.09 | <0.0125 | n.a. | 0.35 | 0.14 |
| Pb (ppm) | 13 | 1 | 9.2 | 0.6 | 9.9 | 0.3 | 0.63 | 0.02 | 3.3 | 0.4 | 6.0 | 0.2 | 5.9 | 0.3 |
| Th (ppm) | 0.32 | 0.11 | 0.14 | 0.06 | 0.54 | 0.10 | 0.66 | 0.35 | 0.45 | 0.11 | 0.10 | 0.03 | 0.23 | 0.03 |
| U (ppm) | 0.37 | 0.01 | 0.17 | 0.07 | 0.31 | 0.08 | 0.48 | 0.34 | 0.20 | 0.07 | 0.05 | 0.004 | 0.61 | 0.06 |

note: Number analysis for major elements in each sample: T2 = 19; T8 = 16; T6 = 14; T9 = 14; T10 = 11. Number analysis for trace elements in each sample: T2 = 5; T8 = 2; T6 = 3; T9 = 4; T10 = 2. σ is standard deviation of the mean value. n. a. = not available.

Table 5. Average chemical compositions of trachyic glasses. Major and trace elements were measured using Electron Microprobe and Laser Ablation ICP-MS respectively.

|  | Short duration experiments | | | | | | | | | | Long duration experiments | | | |
|---|---|---|---|---|---|---|---|---|---|---|---|---|---|---|
|  | T2 | σ | T8 | σ | T6 | σ | T9 | σ | T10 | σ | T11 | σ | T12 | σ |
| $SiO_2$ (wt.%) | 58.25 | 0.44 | 56.93 | 0.57 | 56.43 | 0.32 | 57.96 | 0.51 | 57.93 | 0.38 | 54.46 | 0.65 | 57.49 | 0.80 |
| $TiO_2$ (wt.%) | 0.46 | 0.05 | 0.58 | 0.04 | 0.60 | 0.03 | 0.50 | 0.05 | 0.47 | 0.03 | 0.74 | 0.15 | 0.51 | 0.08 |
| $Al_2O_3$ (wt%) | 17.12 | 0.12 | 16.56 | 0.31 | 16.43 | 0.21 | 16.87 | 0.29 | 16.74 | 0.16 | 17.80 | 0.36 | 13.34 | 0.33 |
| FeO (wt.%) | 2.80 | 0.28 | 3.78 | 0.56 | 4.05 | 0.23 | 3.14 | 0.36 | 3.28 | 0.13 | 3.38 | 0.13 | 3.56 | 0.18 |
| MnO (wt.%) | 0.26 | 0.04 | 0.35 | 0.06 | 0.43 | 0.03 | 0.30 | 0.05 | 0.29 | 0.04 | 0.34 | 0.03 | 0.41 | 0.03 |
| MgO (wt.%) | 0.37 | 0.06 | 0.52 | 0.03 | 0.35 | 0.04 | 0.38 | 0.03 | 0.43 | 0.04 | 0.28 | 0.06 | 0.76 | 0.02 |
| CaO (wt.%) | 1.57 | 0.09 | 1.77 | 0.12 | 1.88 | 0.07 | 1.79 | 0.16 | 1.71 | 0.10 | 1.42 | 0.03 | 1.74 | 0.06 |
| $Na_2O$ (wt.%) | 6.99 | 0.11 | 7.20 | 0.09 | 7.28 | 0.14 | 6.63 | 0.07 | 6.64 | 0.09 | 10.34 | 0.90 | 10.90 | 0.28 |
| $K_2O$ (wt.%) | 6.79 | 0.05 | 6.78 | 0.12 | 6.48 | 0.08 | 6.79 | 0.08 | 6.97 | 0.09 | 5.07 | 0.19 | 5.02 | 0.19 |
| Cl (wt.%) | 0.77 | 0.10 | 1.01 | 0.12 | 1.10 | 0.05 | 0.90 | 0.09 | 0.86 | 0.07 | 1.36 | 0.05 | 1.22 | 0.05 |
| Total | 95.40 | 0.28 | 95.47 | 0.30 | 95.04 | 0.50 | 95.26 | 0.31 | 95.32 | 0.17 | 95.20 | 0.98 | 94.95 | 0.57 |
| Na (ppm) | 51881 | 818 | 53377 | 662 | 54000 | 1067 | 49163 | 547 | 49274 | 672 | 76731 | 6678 | 80831 | 2088 |
| K (ppm) | 56358 | 430 | 56301 | 1035 | 53785 | 655 | 56350 | 701 | 57878 | 769 | 42071 | 1594 | 41709 | 1569 |
| Ca (ppm) | 22457 | 1233 | 25271 | 1732 | 26815 | 1062 | 25615 | 2306 | 24485 | 1399 | 20297 | 429 | 24875 | 787 |
| Li (ppm) | 254 | 3 | 221 | 11 | 247 | 17 | 176 | 5 | 187 | 4 | 62 | 2 | 105 | 5 |
| Be (ppm) | 199 | 62 | 175 | 51 | 128 | 7 | 130 | 20 | 207 | 10 | 266 | 54 | 42 | 9 |
| Sc (ppm) | 133 | 54 | 138 | 51 | 58 | 16 | 87 | 9 | 169 | 18 | 53 | 17 | 10 | 1 |
| V (ppm) | 150 | 60 | 131 | 40 | 80 | 18 | 103 | 4 | 180 | 25 | 102 | 22 | 13 | 2 |
| Co (ppm) | 64 | 14 | 82 | 5 | 38 | 4 | 41 | 6 | 78 | 5 | 40 | 9 | 1.0 | 0.6 |
| Zn (ppm) | 123 | 21 | 163 | 38 | 191 | 51 | 80 | 27 | 132 | 9 | 151 | 29 | 76 | 5 |
| Ga (ppm) | 100 | 10 | 110 | 48 | 58 | 3 | 86 | 2 | 157 | 20 | 155 | 15 | 26 | 6 |
| Ge (ppm) | 52 | 15 | 55 | 24 | 33 | 4 | 44 | 10 | 98 | 9 | 97 | 7 | 7.0 | 0.2 |
| Rb (ppm) | 656 | 14 | 624 | 21 | 657 | 27 | 573 | 4 | 558 | 8 | 500 | 25 | 518 | 3 |
| Sr (ppm) | 16 | 4 | 19 | 4 | 12 | 3 | 16 | 1 | 15 | 5 | 37 | 3 | 14 | 0 |
| Y (ppm) | 192 | 71 | 196 | 44 | 161 | 11 | 143 | 14 | 206 | 8 | 271 | 55 | 110 | 34 |
| Zr (ppm) | 724 | 54 | 960 | 141 | 1091 | 78 | 731 | 93 | 886 | 55 | 1596 | 395 | 1644 | 580 |
| Nb (ppm) | 388 | 148 | 434 | 131 | 381 | 65 | 341 | 37 | 488 | 18 | 731 | 185 | 223 | 87 |
| Cs (ppm) | 479 | 56 | 327 | 32 | 342 | 22 | 267 | 15 | 310 | 11 | 330 | 21 | 62 | 2 |
| Ba (ppm) | 8.9 | 1.9 | 12 | 2 | 6.3 | 2.0 | 8.9 | 1.1 | 7.6 | 2.0 | 18 | 4 | 10 | 1 |
| La (ppm) | 262 | 54 | 285 | 55 | 295 | 9 | 234 | 25 | 308 | 11 | 389 | 53 | 261 | 69 |
| Ce (ppm) | 379 | 78 | 450 | 68 | 504 | 13 | 368 | 41 | 437 | 19 | 566 | 75 | 490 | 131 |
| Pr (ppm) | 127 | 35 | 130 | 34 | 105 | 11 | 98 | 9 | 158 | 5 | 165 | 21 | 49 | 14 |
| Nd (ppm) | 193 | 41 | 214 | 44 | 211 | 8 | 166 | 15 | 233 | 8 | 257 | 28 | 164 | 45 |
| Sm (ppm) | 136 | 45 | 135 | 38 | 102 | 9 | 101 | 8 | 171 | 5 | 170 | 22 | 28 | 8 |
| Eu (ppm) | 79 | 6 | 73 | 8 | 38 | 2 | 55 | 2 | 87 | 1 | 93 | 6 | 2.0 | 0.5 |
| Gd (ppm) | 147 | 13 | 124 | 22 | 97 | 8 | 89 | 10 | 161 | 5 | 154 | 24 | 23 | 7 |
| Tb (ppm) | 137 | 23 | 112 | 25 | 72 | 11 | 74 | 7 | 142 | 4 | 138 | 23 | 3.2 | 1.0 |
| Dy (ppm) | 164 | 20 | 146 | 30 | 98 | 10 | 98 | 9 | 176 | 8 | 197 | 37 | 20 | 6 |
| Er (ppm) | 163 | 30 | 160 | 17 | 88 | 15 | 82 | 12 | 152 | 9 | 168 | 24 | 11 | 3 |
| Yb (ppm) | 180 | 24 | 159 | 42 | 99 | 4 | 95 | 7 | 182 | 10 | 218 | 46 | 11 | 4 |
| Lu (ppm) | 166 | 34 | 148 | 41 | 90 | 12 | 88 | 7 | 177 | 14 | 200 | 48 | 1.6 | 0.5 |
| Hf (ppm) | 173 | 4 | 246 | 61 | 109 | 12 | 79 | 4 | 127 | 23 | 174 | 33 | 34 | 14 |
| Ta (ppm) | 5.7 | 0.5 | 8.0 | 1.0 | 9.1 | 0.5 | 6.0 | 1.1 | 6.9 | 0.6 | 13 | 3 | 10 | 5 |
| Pb (ppm) | 29 | 15 | 16 | 7 | 43 | 6 | 3.0 | 0.2 | 24 | 4 | 37 | 3 | 12 | 1 |
| Th (ppm) | 83 | 17 | 103 | 21 | 106 | 5 | 77 | 11 | 108 | 6 | 165 | 36 | 125 | 41 |
| U (ppm) | 49 | 17 | 56 | 16 | 52 | 4 | 42 | 5 | 67 | 4 | 95 | 23 | 44 | 15 |

note: Number analysis for major elements in each sample: T2 = 9; T8 = 10; T6 = 10; T9 = 10; T10 = 10. Number analysis for trace elements in each sample: T2 = 4; T8 = 7; T6 = 5; T9 = 3; T10 = 4. σ is standard deviation of the mean value.

**Table 6**
Click here to download Table: Table_6.docx

Table 6. Trace elements partition coefficient ($D_i$) between alkali feldspar and trachytic melt.

| | Short duration experiments | | | | | | | | | | Long duration experiments | | | |
|---|---|---|---|---|---|---|---|---|---|---|---|---|---|---|
| | T2 | | T8 | | T6 | | T9 | | T10 | | T11 | | T12 | |
| | $D_i$ | $\sigma D_i$ | $D_i$ | $\sigma D_i$ | $D_i$ | $\sigma D_i$ | $D_i$ | $\sigma D_i$ | $D_i$ | $\sigma D_i$ | $D_i$ | $\sigma D_i$ | $D_i$ | $\sigma D_i$ |
| Na | 0.714 | 0.035 | 0.728 | 0.054 | 0.688 | 0.046 | 0.672 | 0.027 | 0.669 | 0.048 | 0.727 | 0.075 | 0.521 | 0.052 |
| K | 1.313 | 0.085 | 1.194 | 0.123 | 1.360 | 0.110 | 1.451 | 0.070 | 1.449 | 0.090 | 1.104 | 0.129 | 1.637 | 0.200 |
| Ca | 0.951 | 0.247 | 0.945 | 0.187 | 0.786 | 0.185 | 0.665 | 0.192 | 0.603 | 0.141 | 0.246 | 0.034 | 0.451 | 0.089 |
| Li | 0.054 | 0.002 | 0.054 | 0.011 | 0.057 | 0.017 | 0.049 | 0.002 | 0.097 | 0.054 | 0.024 | 0.007 | 0.039 | 0.004 |
| Be | 0.057 | 0.033 | 0.079 | 0.034 | 0.131 | 0.034 | 0.031 | 0.017 | 0.056 | 0.036 | 0.055 | 0.031 | 0.030 | 0.013 |
| Sc | 0.042 | 0.017 | 0.045 | 0.026 | 0.074 | 0.025 | 0.023 | 0.003 | 0.020 | 0.004 | 0.139 | 0.052 | 1.142 | 0.105 |
| V | 0.076 | 0.034 | 0.071 | 0.055 | 0.154 | 0.058 | 0.051 | 0.002 | 0.051 | 0.025 | 0.090 | 0.031 | n.d. | n.d. |
| Co | 0.052 | 0.022 | 0.009 | 0.003 | 0.018 | 0.007 | 0.013 | 0.005 | 0.014 | 0.004 | 0.027 | 0.016 | n.d. | n.d. |
| Zn | 0.055 | 0.015 | 0.022 | 0.014 | 0.014 | 0.008 | 0.028 | 0.019 | 0.016 | 0.006 | 0.017 | 0.003 | n.d. | n.d. |
| Ga | 0.698 | 0.085 | 0.944 | 0.419 | 0.860 | 0.185 | 0.734 | 0.048 | 0.721 | 0.090 | 0.596 | 0.099 | 0.559 | 0.135 |
| Ge | 0.957 | 0.343 | 0.958 | 0.613 | 0.197 | 0.081 | 0.668 | 0.146 | 0.720 | 0.215 | 0.661 | 0.074 | 0.906 | 0.347 |
| Rb | 0.402 | 0.073 | 0.333 | 0.070 | 0.408 | 0.058 | 0.459 | 0.012 | 0.409 | 0.006 | 0.213 | 0.074 | 0.470 | 0.010 |
| Sr | 25.365 | 8.463 | 16.453 | 4.222 | 22.179 | 6.597 | 25.117 | 3.682 | 29.915 | 9.574 | 4.063 | 1.477 | 9.577 | 0.904 |
| Y | 0.002 | 0.001 | 0.002 | 0.001 | 0.004 | 0.0004 | 0.004 | 0.002 | 0.005 | 0.002 | n.d. | n.d. | n.d. | n.d. |
| Zr | 0.008 | 0.002 | 0.008 | 0.001 | 0.045 | 0.007 | 0.020 | 0.004 | 0.012 | 0.002 | 0.001 | 0.0004 | 0.021 | 0.007 |
| Nb | 0.008 | 0.003 | 0.004 | 0.001 | 0.012 | 0.002 | 0.013 | 0.003 | 0.006 | 0.001 | 0.001 | 0.0003 | 0.019 | 0.008 |
| Cs | 0.018 | 0.003 | 0.016 | 0.002 | 0.022 | 0.002 | 0.043 | 0.004 | 0.030 | 0.002 | 0.005 | 0.001 | 0.029 | 0.005 |
| Ba | 55.473 | 21.044 | 27.103 | 5.540 | 45.295 | 17.580 | 56.179 | 8.200 | 64.475 | 17.485 | 7.794 | 2.149 | 15.625 | 1.284 |
| La | 0.132 | 0.029 | 0.111 | 0.027 | 0.110 | 0.007 | 0.097 | 0.011 | 0.135 | 0.041 | 0.014 | 0.005 | 0.066 | 0.018 |
| Ce | 0.077 | 0.022 | 0.057 | 0.012 | 0.058 | 0.007 | 0.046 | 0.008 | 0.074 | 0.029 | 0.009 | 0.002 | 0.039 | 0.011 |
| Pr | 0.050 | 0.016 | 0.041 | 0.011 | 0.047 | 0.010 | 0.023 | 0.002 | 0.040 | 0.022 | 0.006 | 0.002 | 0.030 | 0.010 |
| Nd | 0.040 | 0.014 | 0.034 | 0.008 | 0.029 | 0.012 | 0.021 | 0.009 | 0.021 | 0.011 | 0.005 | 0.002 | 0.031 | 0.009 |
| Sm | 0.029 | 0.010 | 0.018 | 0.008 | 0.023 | 0.009 | 0.006 | 0.003 | 0.016 | 0.010 | n.d. | n.d. | n.d. | n.d. |
| Eu | 1.438 | 0.282 | 0.754 | 0.112 | 1.427 | 0.131 | 1.098 | 0.074 | 0.854 | 0.094 | 0.453 | 0.060 | 0.818 | 0.210 |
| Gd | 0.004 | 0.002 | 0.005 | 0.002 | 0.011 | 0.002 | 0.010 | 0.005 | 0.007 | 0.000 | n.d. | n.d. | n.d. | n.d. |
| Tb | 0.004 | 0.001 | 0.004 | 0.001 | 0.005 | 0.001 | 0.005 | 0.001 | 0.003 | 0.001 | 0.001 | 0.0002 | n.d. | n.d. |
| Dy | n.d. | n.d. | 0.005 | 0.002 | 0.008 | 0.003 | 0.005 | 0.001 | 0.004 | 0.002 | n.d. | n.d. | n.d. | n.d. |
| Er | 0.003 | 0.001 | 0.002 | 0.001 | 0.005 | 0.001 | 0.006 | 0.002 | 0.005 | 0.002 | n.d. | n.d. | n.d. | n.d. |
| Yb | n.d. | n.d. | 0.005 | 0.002 | 0.006 | 0.003 | 0.005 | 0.002 | 0.005 | 0.002 | n.d. | n.d. | n.d. | n.d. |
| Lu | 0.004 | 0.001 | 0.002 | 0.001 | 0.006 | 0.001 | 0.005 | 0.002 | 0.004 | 0.002 | n.d. | n.d. | n.d. | n.d. |
| Hf | 0.015 | 0.003 | 0.033 | 0.015 | 0.078 | 0.015 | 0.024 | 0.005 | 0.020 | 0.007 | 0.001 | 0.0004 | n.d. | n.d. |
| Ta | 0.128 | 0.048 | 0.050 | 0.019 | 0.054 | 0.005 | 0.017 | 0.006 | 0.029 | 0.013 | n.d. | n.d. | 0.033 | 0.021 |
| Pb | 0.443 | 0.232 | 0.564 | 0.257 | 0.231 | 0.032 | 0.213 | 0.017 | 0.137 | 0.028 | 0.161 | 0.014 | 0.477 | 0.035 |
| Th | 0.004 | 0.002 | 0.001 | 0.001 | 0.005 | 0.001 | 0.009 | 0.005 | 0.004 | 0.001 | 0.001 | 0.0002 | 0.002 | 0.001 |
| U | 0.008 | 0.003 | 0.003 | 0.001 | 0.006 | 0.002 | 0.012 | 0.008 | 0.003 | 0.001 | 0.001 | 0.0001 | 0.014 | 0.005 |

note: $D_i$ = average partition coefficient. $\sigma D_i$ = standard deviation of $D_i$. n. d. = $D_i$ value does not determined because under limit detection.

Table 7. Fit parameters from Lattice Strain Model (LSM).

|  | Short duration experiments | | | | | Long duration experiments | |
|---|---|---|---|---|---|---|---|
|  | T2 | T8 | T6 | T9 | T10 | T11 | T12 |
| 1+ Valence | | | | | | | |
| $D_0$ | 2.95 (0.08) | 2.47 (0.32) | 2.97 (0.22) | 2.97 (0.35) | 3.30 (0.15) | 3.76 (0.09) | 3.42 (0.21) |
| $E_0$ | 81 (2) | 75 (9) | 80 (5) | 79 (8) | 88 (3) | 105 (2) | 92 (4) |
| $r_0$ | 1.462 | 1.465 | 1.464 (0.002) | 1.468 (0.003) | 1.466 (0.01) | 1.452 | 1.476 (0.002) |
| 2+ Valence | | | | | | | |
| $D_0$ | 64.03 (1.22) | 36.49 (0.44) | 63.52 (5.45) | 70.15 (0.16) | 87.51 (3.16) | 10.7 (0.73) | 21.76 (3.98) |
| $E_0$ | 125 (4) | 143 (3) | 168 (22) | 153 (1) | 179 (12) | 158 (17) | 145 (47) |
| $E_c$ | 100 | 101 | 100 | 99 | 98 | 104 | 101 |
| $r_0$ | 1.476 (0.001) | 1.460 | 1.460 | 1.470 | 1.467 | 1.460 | 1.452 |
| $r_c^{2+}$ | 1.445 | 1.437 | 1.444 | 1.457 | 1.459 | 1.408 | 1.438 |

note: all calculations are related to the alkali-feldspar A-site. $D_0$: strain-compensated partition coefficient. $E_0$: Young Modulus of the lattice site (GPa). $r_0$: optimum site radius (Å). $E_c$: Young Modulus (GPa) calculated following Blundy and Wood (2003). $r_c$: site radius (Å) calculated following Blundy and Wood (2003).

Table 8. Selected experimental alkali feldspar/liquid partition coefficients for mono- and divalent cations.

| Reference | Ice96 | Ice96 | Ice96 | Ice96 | Ice96 | Ice96 | Ice96 | Ice96 | Ice96 | Ice96 | Ice96 | Ice96 | Ice96 | Ice96 | Ice96 |
|---|---|---|---|---|---|---|---|---|---|---|---|---|---|---|---|
| Sample | 5+15 | 5+14 | 5+8 | 5+7 | 5+5 | 5+9 | 6+6 | 6+5 | 6+7 | 6+4 | 7+6 | 7+7 | 7+4 | | |
| $P$, GPa/$T$ °C | 0.2/650 | 0.2/650 | 0.2/650 | 0.2/650 | 0.2/700 | 0.2/700 | 0.2/650 | 0.2/700 | 0.2/700 | 0.2/750 | 0.2/650 | 0.2/750 | 0.2/700 | | |
| Composition | metapel | metapel | metapel | metapel | metapel | metapel | metapel | metapel | metapel | metapel | metapel | metapel | metapel | | |
| Analytical technique | EMPA | EMPA | EMPA | EMPA | EMPA | EMPA | EMPA | EMPA | EMPA | EMPA | EMPA | EMPA | EMPA | | |
| Time (days) | 28 | 28 | 28 | 28 | 28 | 28 | 28 | 28 | 28 | 28 | 28 | 28 | 28 | | |
| Cs | 0.11 | 0.10 | 0.10 | 0.13 | 0.11 | 0.12 | 0.10 | 0.12 | 0.12 | 0.13 | 0.09 | 0.10 | 0.12 | | |
| K | 0.55 | 0.88 | 0.48 | 1.85 | 1.93 | 0.45 | 0.40 | 0.63 | 0.37 | 0.68 | 0.37 | 0.30 | 1.68 | | |
| Na | 3.16 | 3.75 | 2.88 | 1.29 | 1.13 | 2.55 | 2.63 | 2.31 | 2.29 | 1.84 | 3.13 | 2.45 | 0.96 | | |
| Rb | 0.08 | 0.09 | 0.08 | 0.45 | 0.38 | 0.09 | 0.07 | 0.08 | 0.08 | 0.08 | 0.08 | 0.08 | 0.31 | | |
| Ba | 1.26 | 2.43 | 1.18 | 11.03 | 12.59 | 0.31 | 4.55 | 2.61 | 6.20 | 5.42 | 1.07 | 2.89 | 12.93 | | |
| Sr | 7.96 | 18.38 | 15.00 | 10.51 | 13.98 | 13.41 | 17.91 | 9.16 | 23.05 | 12.44 | 10.95 | 12.64 | 10.09 | | |
| Ca | 4.95 | 4.35 | 5.53 | 0.96 | 0.65 | 5.42 | 8.70 | 6.97 | 5.10 | 3.99 | 6.58 | 6.98 | 0.65 | | |
| Mg | 0.56 | 0.82 | 0.75 | 0.69 | 1.29 | 1.00 | 1.80 | 1.29 | 1.29 | 0.75 | 1.13 | 1.13 | 0.90 | | |

Table x. continued

| Reference | He12 | He12 | He12 | He12 | He12 | He12 | He12 | He12 | He12 | He12 | He12 | Mo03 | Mo03 | Mo03 | Fab09 | Fab09 | He12 | He12 | He12 | He12 | He12 | He12 | He12 | He12 |
|---|---|---|---|---|---|---|---|---|---|---|---|---|---|---|---|---|---|---|---|---|---|---|---|---|
| Sample | 203 | 204 | 236r | 304 | 254 | 414 | 436 | 431 | 511 | 484 | 341 | 373 | 351 | BC2-9 | BC2-14 | Run 7 | Run 11 | 256 | 231 | 238 | 301 | 234 | 402 | |
| $P$, MPa/$T$°C | 0.1/840 | 0.1/855 | 0.1/825 | 0.1/835 | 0.1/815 | 0.1/790 | 0.1/760 | 0.1/760 | 0.1/760 | 0.1/740 | 0.1/905 | 0.1/905 | 0.1/850 | 0.2/750 | 0.2/750 | 0.8/750 | 0.8/750 | 0.1/815 | 0.1/825 | 0.1/810 | 0.1/845 | 0.1/825 | 0.1/875 | |
| Composition | trachyte | trachyte | trachyte | trachyte | trachyte | trachyte | trachyte | trachyte | trachyte | trachyte | trachyte | trachyte | trachyte | rhyolite | rhyolite | trachyte | trachyte | trachyte | trachyte | trachyte | trachyte | trachyte | trachyte | |
| Analytical technique | EMPA | EMPA | EMPA | EMPA | EMPA | EMPA | EMPA | EMPA | EMPA | EMPA | EMPA | EMPA | EMPA | EMPA | EMPA | LA | LA | EMPA | EMPA | EMPA | EMPA | EMPA | EMPA | |
| Time (days) | - | - | - | - | - | - | - | - | - | - | - | - | 5 | 14 | 3.8 | 5 | - | - | - | - | - | - | | |
| Cs | 0.35 | 0.36 | 0.32 | 0.33 | 0.22 | 0.34 | 0.34 | 0.35 | 0.30 | 0.31 | 0.45 | 0.40 | 0.35 | 0.063 | 0.063 | 0.21 | | 0.55 | 0.49 | 0.37 | 0.43 | 0.36 | 0.39 | |
| K | 0.97 | 0.87 | 0.81 | 0.79 | 0.83 | 0.87 | 0.88 | 0.76 | 0.71 | 0.69 | 1.16 | 1.06 | 0.85 | 2.00 | 2.09 | 2.43 | 2.32 | 1.19 | 1.19 | 1.16 | 1.60 | 1.61 | 1.00 | |
| Na | 0.04 | 0.07 | 0.07 | 0.06 | 0.08 | 0.05 | 0.04 | 0.11 | 0.14 | 0.18 | 0.08 | 0.09 | 0.06 | 0.64 | 0.63 | 0.82 | 0.62 | 0.71 | 0.68 | 0.67 | 0.73 | 0.52 | 0.06 | |
| Rb | | | | | | | | | | | | | | | | 1.03 | 1.00 | 0.40 | 0.46 | 0.41 | 0.21 | 0.47 | | |
| Ba | | | | | | | | | | | | | | | | 4.61 | 10.26 | 55.5 | 56.2 | 64.5 | 7.79 | 15.6 | | |
| Sr | | | | | | | | | | | 9.7 | | | | | 4.28 | 4.11 | 25.4 | 25.1 | 29.9 | 4.06 | 9.58 | | |
| Ca | | | | | | | | | | | 10.37 | | | | | 0.60 | 0.23 | 0.95 | 0.67 | 0.60 | 0.25 | 0.45 | | |
| Mg | | | | | | | | | | | 0.55 | | | | | 0.22 | 0.045 | 0.081 | 0.079 | 0.070 | 0.11 | 0.039 | | |

Note: Ice96: Icenhower and London (1996), Mo03: Morgan and London (2003), Fab09: Fabbrizio et al. (2009), He12: Henderson and Pierozynski (2012). TS: this work. Metapel: metapelite. Phonol: phonolite. LA: LA-ICP-MS.

Table 9. Selected partition coefficients alkali-feldspar/liquid for natural trachytic samples.

| Reference | Villemant 1988 | | | | White et al. 2003 | | | | | | JWS-PS | Pappalardo et al. 2008 Trachyte | Fedele et al. 2015 | |
|---|---|---|---|---|---|---|---|---|---|---|---|---|---|---|
| Sample | CFP8 | CFA31 | ZR2B | CFP2 | 98520 | 98521 | 98522 | 98523 | 98626 | 98527 | 98531 | | AI(T) | AN(T) |
| Be | | | | | | 0.08 | | 0.08 | | | | | | |
| Na | | | | | | | | | | | | | 0.44 | 0.45 |
| K | | | | | | | | | | | | | 1.55 | 1.53 |
| Ca | | | | | | | | | | | | | 0.21 | 0.20 |
| Sc | 0.04 | 0.07 | 0.03 | 0.05 | | 0.04 | | | 0.02 | | 0.02 | | 0.43 | 0.34 |
| Co | 0.02 | 0.04 | 0.03 | 0.02 | | 1.09 | | | 1.18 | | 1.16 | | | |
| Zn | | | | | | | | | | | | | | |
| Ga | | | | | | | | | | | | | | |
| Rb | 0.9 | 0.97 | 0.8 | 0.7 | 0.26 | 0.21 | 0.22 | 0.31 | 0.28 | | 0.25 | | 0.67 | 0.76 |
| Sr | 2 | 3.3 | 3.4 | 6.6 | 2.32 | 3.29 | 2.62 | | 2.03 | 1.51 | 4.06 | 2 | 2.94 | 2.26 |
| Y | | | | | | 0.02 | 0.02 | | 0.01 | 0.03 | 0.02 | | 0.001 | |
| Zr | 0.17 | 0.12 | 0.12 | 0.08 | 0.03 | 0.02 | 0.03 | 0.03 | 0.02 | 0.04 | 0.04 | | | |
| Nb | | | | | | 0.02 | | | 0.02 | | 0.02 | | | |
| Cs | 0.11 | 0.12 | 0.08 | 0.12 | | | | | | | | | 0.06 | 0.07 |
| Ba | 1.13 | 5.89 | 6.5 | 16.6 | 5 | 10.63 | 9.41 | 3.83 | 3.13 | 1.87 | 12.22 | 4.04 | 7.01 | 2.81 |
| La | 0.1 | 0.09 | 0.09 | 0.08 | | 0.02 | | | 0.01 | | 0.02 | 13 | 0.04 | 0.08 |
| Ce | | | | | | 0.02 | 0.01 | | 0.01 | | 0.02 | 0.13 | 0.02 | 0.03 |
| Pr | | | | | | 0.02 | | | 0.02 | | 0.02 | 0.04 | 0.01 | 0.02 |
| Nd | | | | | | 0.02 | | | 0.02 | | 0.02 | | 0.004 | 0.01 |
| Sm | | | | | | 0.02 | | | 0.02 | | 0.02 | 0.01 | 0.01 | |
| Eu | 0.97 | 1.06 | 0.94 | 0.85 | | 0.05 | | | 0.18 | | 0.24 | 0.9 | 0.74 | 1.04 |
| Gd | | | | | | 0.02 | | | 0.01 | | 0.02 | | 0.02 | 0.03 |
| Tb | 0.03 | 0.02 | 0.3 | 0.03 | | | | | | | | | 0.01 | 0.02 |
| Dy | | | | | | 0.02 | | | 0.01 | | 0.02 | | 0.01 | 0.03 |
| Er | | | | | | 0.02 | | | 0.01 | | 0.02 | | 0.01 | |
| Yb | | | | | | 0.02 | | | 0.01 | | 0.03 | | | |
| Hf | 0.38 | 0.03 | 0.02 | 0.13 | | | | | | | | | | |
| Ta | 0.01 | | 0.02 | 0.01 | | | | | | | | | 0.01 | 0.57 |
| Pb | | | | | | | | | | | | | 0.38 | |
| Th | 0.02 | 0.03 | 0.02 | 0.01 | | | | | | | | | | |
| U | 0.02 | 0.03 | 0.02 | 0.01 | | | | | | | | | | |

Table 10
Click here to download Table: Table_10.docxTable 10. Crystallization kinetics of alkali feldspar in Phlegrean trachytic magmas.

| Sample | | $D_{Ba}$ | $D_{Sr}$ | time $D_{Ba}$ (days)* | time $D_{Sr}$ (days)# |
|---|---|---|---|---|---|
| Fedele et al. (2015) | B | 21 | 8.8 | 6 | 7 |
| Fedele et al. (2015) | R | 20 | 8.7 | 6 | 7 |
| Fedele et al. (2015) | I | 16 | 7.5 | 7 | 8 |
| Fedele et al. (2015) | AI | 2.8 | 2.5 | 10 | 10 |
| Fedele et al. (2015) | AN | 2.5 | 2.8 | 10 | 10 |
| Fedele et al. (2015) | AN | 5.4 | 3.1 | 9 | 9 |
| Villemant (1988) | Feld/Trach | 16.6 | 6.6 | 7 | 8 |
| Villemant (1988) | Feld/Trach | 1.13 | 2 | 10 | 10 |
| Pappalardo et al. (2008) | Ksf/T | 13 | - | 8 | - |
| Pappalardo et al. (2008) | Kfs/T | 0.3 | 2 | 10 | 10 |
| Pappalardo et al. (2008) | Ksf/Tp | 7.7 | 7.1 | 9 | 8 |

note: partition coefficients of Ba and Sr between alkali feldspar and trachytic melt, calculated from Villemant (1998), Pappalardo et al. (2008) and Fedele et al. (2015), are used to estimate the residence time of Phlegrean magma through Eq. 5* and Eq. 6#.

**Supplementary Table 1**

[Click here to download Background dataset for online publication only: Supplementary Table 1.docx]